\PassOptionsToPackage{table}{xcolor}
\documentclass[manuscript, screen, nonacm]{acmart}
\usepackage[utf8]{inputenc}
\usepackage{xspace}
\usepackage{ifthen}
\usepackage[T1]{fontenc}
\usepackage{amsmath}
\usepackage{graphicx}
\usepackage{svg}
\usepackage{listings}
\usepackage{xcolor}
\usepackage{caption}
\usepackage{balance}
\usepackage{subcaption}
\usepackage{listings}
\usepackage{tabularx}
\usepackage{booktabs, multirow} 
\usepackage{algorithm}
\usepackage{algpseudocode}
\usepackage{numprint}
\usepackage[most]{tcolorbox}
\usepackage{url}
\usepackage{tikz}
\usepackage{makecell}
\usepackage[framemethod=tikz]{mdframed}
\mdfdefinestyle{mpdframe}{
    frametitlebackgroundcolor   =black!15,
    frametitlerule              =true,
    roundcorner                 =1pt,
    middlelinewidth             =1pt,
    skipabove                   =\baselineskip,
    innermargin                 =0.1cm,
    innerleftmargin             =0.1cm,
    innerrightmargin            =0.1cm,
    innertopmargin              =0.1cm,
    innerbottommargin           =0.1cm
}

\definecolor{mGreen}{rgb}{0,0.6,0}
\definecolor{mGray}{rgb}{0.5,0.5,0.5}
\definecolor{mPurple}{rgb}{0.58,0,0.82}
\definecolor{backgroundColour}{rgb}{0.95,0.95,0.92}

\lstdefinestyle{CStyle}{
    basicstyle=\scriptsize\ttfamily,
    commentstyle=\color{mGreen},
    keywordstyle=\color{blue},
    numberstyle=\tiny\color{mGray},
    stringstyle=\color{mPurple},
    breakatwhitespace=false,         
    breaklines=true,                 
    captionpos=b,                    
    keepspaces=true,                 
    numbers=left,                    
    numbersep=5pt,                  
    showspaces=false,                
    showstringspaces=false,
    showtabs=false,                  
    tabsize=2,
    language=C,
    frame=single
}

\lstdefinestyle{GoStyle}{
    basicstyle=\scriptsize\ttfamily,
    commentstyle=\color{mGreen},
    keywordstyle=\color{blue},
    numberstyle=\tiny\color{mGray},
    stringstyle=\color{mPurple},
    breakatwhitespace=false,         
    breaklines=true,                 
    captionpos=b,                    
    keepspaces=true,                 
    numbers=left,                    
    numbersep=5pt,                  
    showspaces=false,                
    showstringspaces=false,
    showtabs=false,                  
    tabsize=2,
    language=go,
    frame=single
}

\lstdefinestyle{bash}{
    basicstyle=\scriptsize\ttfamily,
    commentstyle=\color{mGreen},
    keywordstyle=\color{blue},
    numberstyle=\tiny\color{mGray},
    stringstyle=\color{mPurple},
    breakatwhitespace=false,         
    breaklines=true,                 
    captionpos=b,                    
    keepspaces=true,                
    showspaces=false,                
    showstringspaces=false,
    showtabs=false,                  
    tabsize=2,
    language=go,
    frame=single
}

\title{Galápagos: Automated N-Version Programming with LLMs}

\author{Javier Ron}
\email{javierro@kth.se}
\affiliation{%
  \institution{KTH Royal Institute of Technology}
  \city{Stockholm}
  \country{Sweden}
}

\author{Diogo Gaspar}
\email{dgaspar@kth.se}
\affiliation{%
  \institution{KTH Royal Institute of Technology}
  \city{Stockholm}
  \country{Sweden}
}

\author{Javier Cabrera-Arteaga}
\email{javierca@kth.se}
\affiliation{%
  \institution{KTH Royal Institute of Technology}
  \city{Stockholm}
  \country{Sweden}
}

\author{Benoit Baudry}
\email{benoit.baudry@umontreal.ca}
\affiliation{%
  \institution{Université de Montréal}
  \city{Montreal}
  \country{Canada}
}

\author{Martin Monperrus}
\email{monperrus@kth.se}
\affiliation{%
  \institution{KTH Royal Institute of Technology}
  \city{Stockholm}
  \country{Sweden}
}

\newcommand{\revised}[1]{{\color{black}#1}}

\newcommand{\toolname}{\textsc{Galápagos}\@\xspace}
\newcommand{\etal}{et al.\@\xspace}

\begin{document}
\begin{abstract}
N-Version Programming is a well-known methodology for developing fault-tolerant systems. 
It achieves fault detection and correction at runtime by adding diverse redundancy into programs, minimizing fault mode overlap between redundant program variants.
In this work, we propose the automated generation of program variants using large language models.
We design, develop and evaluate \toolname:
a tool for generating program variants using LLMs, validating their correctness and equivalence, and using them to assemble N-Version binaries.
We evaluate \toolname by creating N-Version components of real-world C code.
Our original results show that \toolname can produce program variants that are proven to be functionally equivalent, even when the variants are written in a different programming language.
Our systematic diversity measurement indicates that functionally equivalent variants produced by \toolname, are statically different after compilation, and present diverging internal behavior at runtime.
We demonstrate that the variants produced by \toolname can protect C code against real miscompilation bugs which affect the Clang compiler.
Overall, our paper shows that producing N-Version software can be drastically automated by advanced usage of practical formal verification and generative language models.
\end{abstract}

\ccsdesc[500]{Software and its engineering~Software development techniques}

\keywords{N-Version Programming, Large Language Models.}

\maketitle

\section{Introduction}

N-Version programming is a software engineering approach to reliability based on crafting software diversity~\cite{building} with different teams~\cite{avizienis1985n}. It is predominantly  used to enhance the reliability of mission-critical systems~\cite{avizienis-2}.
This is achieved by building diverse and equivalent software components, with the objective that these would not fail all at once, under the same conditions, hence making the whole system more robust~\cite{avizienis-2}.

\revised{However, N-Version programming faces two fundamental challenges:
First, empirical studies have shown that implementations developed by different teams might not be truly independent with respect to the target fault model, as developers tend to share similar mental models and processes, leading to correlated failures~\cite{knight2012experimental,brilliant2002analysis}.
Second, and the focus of this work, is the significant cost overhead of developing multiple versions. 
There are high costs induced by the manual development and verification of multiple program versions instead of a single one, typically linear in the number of versions~\cite{simplicity}.
Another important additional cost is the development of additional software to orchestrate the multiple versions (the N-Version harness).
Our core insight is that LLMs can bring down the cost of engineering N-Version software.}

Large Language Models (LLMs) are powerful tools able to carry out complex source code-related tasks, such as code completion~\cite{codex}, automatic program repair~\cite{cigar}, and code summarization~\cite{ahmed2022few}.
Our key intuition is that they are able to efficiently automate N-Version programming.
Given a reference program, we can leverage the creative and search power of LLMs to automatically produce program variants.
We anticipate these variants to be diverse, due to the great volume of data used in LLM training, apt to being combined in an N-Version system.
In other words, we want to harness the LLM's ability to synthesize program variants, and instead of having multiple teams producing multiple versions at a high costs, having the LLM producing them at virtually no cost.

In this paper, we design, implement and evaluate \toolname, a tool for automated, verified N-Version programming.
\toolname performs three passes: incl. LLM prompting and program equivalence verification to provide strong guarantees.
\toolname generates variants of a reference function and then assembles N-Version implementations of that function.
We implement \toolname to harden C libraries, with support for producing same-language variants (C-to-C), and cross-language variants (C-to-Go).

\begin{figure}
    \centering
    \begin{minipage}{.49\textwidth}
    \begin{lstlisting}[style=CStyle]
//main.c
int printf(const char *, ...);
static int a = -3, b;
static char c;
int d;
int e(int f, int g) {
  if (f - g < 10000)
    return f;
  return f + 1 % -f;
}
int main() {
  int *h[] = {&a, &a};
  for (; c <= 37; ++c) {
    int *i = &b;
    *i |= e(a, 8) + d;
  }
  printf("%d\n", b);
}
    
\end{lstlisting}
\end{minipage}
\hfill
\begin{minipage}{.49\textwidth}
\begin{lstlisting}[style=bash]
$ clang -v
clang version 17.0.0
Target: x86_64-unknown-linux-gnu

...

$ clang -O0 main.c; ./a.out
-3

$ clang -Os main.c; ./a.out
-1

\end{lstlisting}
\end{minipage}
    \caption{C code which triggers a miscompilation on clang v17.0.0, when using the \texttt{-Os} flag (left). After running the compiled binary, it prints \texttt{-1}. The correct result is \texttt{-3} (right). \toolname is able to detect and mitigate this miscompilation fault.}
    \label{fig:miscompilation}
\end{figure}

We conduct a systematic evaluation of \toolname to quantify the correctness, diversity, and usefulness of both the function variants and the N-Version components.
First, we assemble a dataset of 30  real-world C functions, extracted from 6 notable  cryptography and multimedia open source projects. We run \toolname to transform each function in the dataset into an N-Version component.
Second, we measure the diversity of the variants, both statically and dynamically.
Statically, we measure the uniqueness of the resulting machine code after compiling the variants, using different optimization configurations.
\revised{Dynamically, we execute each unique variant, and record the CPU instructions they utilize, comparing against state-of-the-art diversification and obfuscation tool Tigress~\cite{tigress} as a baseline.}
This allows us to quantify the variation in execution paths, which is known to be important against certain types of exploits such as side-channel attacks~\cite{wasm-mutate}.
Third, we rigorously test the assembled N-Version implementations in the context of miscompilation bugs~\cite{yang2011finding}.
It clearly demonstrates \toolname's ability to enhance robustness wrt. our target fault model.

Our experiments demonstrate several key findings:
First, \toolname uses LLMs to generate code variants that are provably equivalent, even when these variants are written in different programming languages.
During our experiments, \numprint{234} \textit{equivalent} variants were found from a dataset of 30 reference programs.
\revised{Second, \toolname is capable of producing and identifying code variants that exhibit diversity both in their code (statically) and during execution (dynamically), outperforming transformation-based diversification techniques like Tigress.}
From the dataset of 30 programs, 126 \textit{unique} variants were verified as different, even after all compilation and optimization passes.
Last but not least, these diverse code variants can be utilized to strengthen critical sections of software via automated N-Version programming: we demonstrate \toolname' capability to mitigate real-world, reported miscompilation bugs in the Clang compiler.

In summary, our contributions are
\begin{itemize}
    \item The design and implementation of \toolname, a tool for automated and verified N-Version programming using LLMs.
    As far as we are aware, this is the first work to realize an automatic N-Version programming framework with formal guarantees. \toolname is a major contribution towards reducing the high costs of N-Version programming.
    \item A systematic suite of experiments to evaluate the correctness, value, and practicality of automated N-Version programming with \toolname. The experiments consist of transforming real-world C code from notable open-source libraries into an N-Version counterpart.
    \item A publicly available, open-source repository for experimental reproduction and future research on automated N-Version programming, at~ \url{https://github.com/ASSERT-KTH/Galapagos/}
\end{itemize}

\section{Background}

In this section, we introduce the two areas in which our work is rooted: N-Version programming and neural machine translation.

\subsection{N-Version Programming}

N-Version programming is a software development approach, which consists of creating $N$ implementations or \textit{versions} of a specific program~\cite{avizienis1985n}.
The core idea behind this approach is that when these versions are simultaneously executed, errors can be timely detected and mitigated by comparing their outputs.
Ideally, the difference in implementations between versions is maximal, such that any coincidental errors are avoided.~\cite{1701974}.
While originally devised as a fault-tolerance mechanism, N-Version programming has been adapted to enhance other specific properties of software, such as availability~\cite{neth}, reliability~\cite{error-recovery, verifiable-client-diversity}, performance~\cite{n-version-perf}, or security~\cite{n-version-sec, back-to-the-future}  

However, the enhancements offered by N-Version programming come with an attached trade-off, as it introduces many challenges throughout the software development lifecycle.
These challenges include increased maintenance overhead, increased compute and memory use, or interoperability issues~\cite{n-version-challenges, neth}.
Addressing these challenges requires additional effort and careful coordination across engineering teams.

\begin{figure*}
    \centering
    \includegraphics[width=\textwidth]{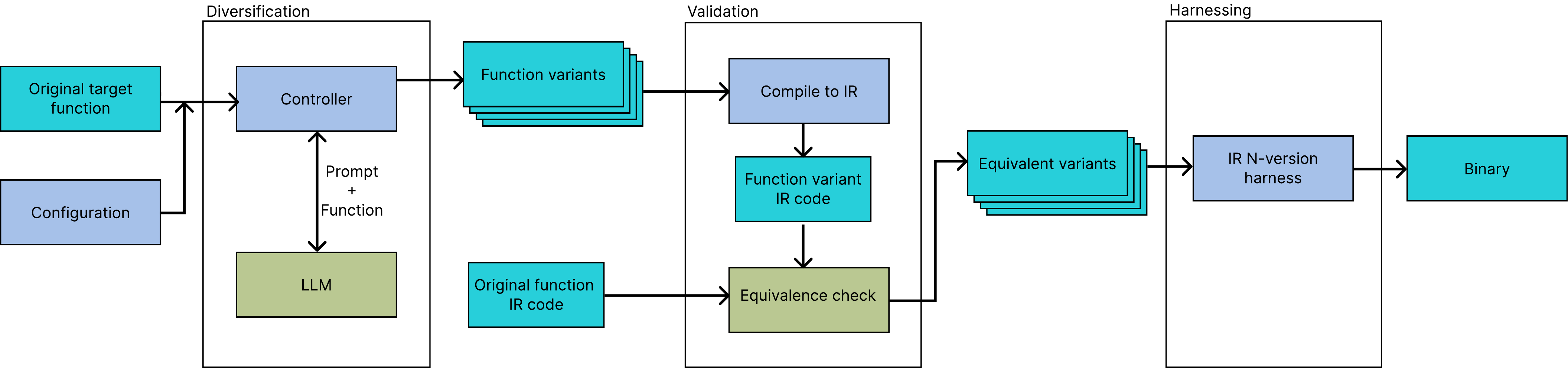}
    \caption{\revised{\toolname pipeline for automated, verified N-Version programming.}}
    \label{fig:pipeline-overview}
\end{figure*}

An essential challenge is the increased cost of development, given that the time and resources needed to develop the versions increase at least linearly with N.
To address this challenge, automating the process for creating new versions is a known and well-studied approach~\cite{cadar-nvx, wasm-mutate, n-version-obf, los-alamos}. In this paper, we contribute to this field of automatic synthesis of diverse program versions.

\subsection{Code Translation with LLMs}
Neural machine translation (NMT) for source code refers to the application of machine learning techniques for the automated translation between programming languages~\cite{tufano-nmt}.
It aims to overcome the challenges associated with traditional rule-based or statistical translation methods by leveraging the power of neural networks to capture and model complex patterns inherent in source code~\cite{sequencer}.
The core principle behind NMT for source code involves training a neural network model on large datasets consisting of source code snippets in different programming languages and their corresponding translations~\cite{facebook-nmt}.
During the training phase, the model learns to encode the source code syntax and semantics into a continuous representation, enabling it to generate accurate translations.
Subsequently, the trained model can be deployed to translate code snippets from one programming language to another automatically.
NMT is tied to recent developments in large language models, which have proven to be efficient in performing translation tasks~\cite{vert}.

\section{N-Version programming with LLMs}
\label{sec:design}

In this section, we present \toolname, a tool that leverages large language models to harden software via automated N-Version programming.
\toolname works at the source-code level for generation, and at the intermediate representation~(IR) level for verification.
At the source-code level, it uses an LLM to produce new, diversified variants of program functions.
At the intermediate representation level, these variants are verified to be semantically equivalent and to be finally assembled into N-Version functions.
Because the variants are formally verified to be equivalent, \toolname is able to provide strong guarantees about the behavior final N-Version assembly, irrespective of the introduced software diversity.

\subsection{Fault Model}
\label{sec:fault-model}
In this paper, we propose a novel technique to mitigate the class of faults called \textit{miscompilation}.
Miscompilations occur when a compiler erroneously generates machine code that deviates from the behavior defined in the source code, due to internal bugs in the compiler itself.
These compiler faults are very hard to identify because they are typically silent. They are critical because the resulting application binaries behave incorrectly at runtime~\cite{sun2016toward}.

Miscompilations are present in all major compilers \cite{compcert}.
As of May 2025, Clang and GCC have \href{https://github.com/llvm/llvm-project/issues?q=is%3Aopen+is%3Aissue+label%3Amiscompilation}{396} and \href{https://gcc.gnu.org/bugzilla/buglist.cgi?quicksearch=miscompilation}{44} unresolved reports of miscompilation bugs, respectively.
As an example, consider the code from an open \href{https://github.com/llvm/llvm-project/issues/68871}{LLVM bug report} in \autoref{fig:miscompilation}.
After compilation using \texttt{clang} v17.0.0, with the \texttt{-Os} optimization flag, the result is a binary which produces incorrect output: \texttt{-1}.
The compilation process exits ``successfully'' with code \texttt{0}, and offers no warnings or error messages which would hint that something has gone wrong.
Hence, there is an error in one of Clang's optimization passes.
This kind of fault can cause major downstream failures \cite{marcozzi2019compiler}.

Let us abstract the compilation process as the following composition of functions:

\vspace{1pt}
\begin{equation}
P = B_{a}(M(F_{l}(S)))
\label{eq:compilation}
\end{equation}
\vspace{1pt}

In \autoref{eq:compilation}, $S$ is a program's source code; $F_{l}$ is the front-end stage, which is specific to the language $l$, and transforms $S$ to an intermediate representation (IR); $M$ is the middle-end phase, i.e. a set of optimization passes performed on the IR; $B_{a}$ is the back-end stage, which transforms the IR into assembly code targeted towards a specific architecture ${a}$; and $P$ is the resulting executable program.
Furthermore, $M$ is subject to a wide range of configuration options, each of which can potentially introduce miscompilation bugs.
For example, optimization flags such as \texttt{-Os} or \texttt{-O3} enable aggressive transformations intended to reduce code size or improve performance, but they also increase the risk of activating miscompilation bugs in these optimization passes.

In this work, we want to harden programs generated by the compilers' middle-end $M$ and back-end $B_{a}$ stages, i.e. any operations performed on the intermediate representation which result in $P$ not behaving as specified by $S$.

\subsection{Overview}
\toolname is a three-pass pipeline, as illustrated in~\autoref{fig:pipeline-overview}: \textit{Diversification}, \textit{Validation}, and \textit{Harnessing}.

The pipeline's input is a function's source code, which is part of a larger application or library.
For instance, a function could compute a cryptographic hash.
Hereafter, we refer to this function as the \textit{reference function} or simply the \textit{reference}.
The reference defines the expected functionality of the function on the whole input domain.
In other words, the reference formally specifies the input/output behavior that all variants should conform to.

To create an N-Version implementation of the reference function, it is first processed in the Diversification pass (\autoref{sec:diversification}) of \toolname.
Here, an LLM is prompted to automatically generate different \textit{variants} of the reference.
Second, the collection of generated variants is passed to the Validation pass (\autoref{sec:validation}), which filters out non-viable variants produced by the Diversification pass. The Validation pass proceeds through a sequence of  validation steps:
(1) \toolname compiles each variant and produces its corresponding IR code, variants that cannot be compiled are filtered out as non-viable;
(2) The IR code of all variants that compile is tested, and sent to a formal equivalence checker to compare their functionality against the reference.
The variants that pass the tests and the equivalence check are then forwarded to the next pass of \toolname.
This ensures that after this pass, the ensuing variants have formal guarantees that the original functionality is preserved, and expected diverse internal behavior per the LLM instructions.
Third, the Harnessing pass (\autoref{sec:harnessing}) uses the resulting variants to assemble an executable, where the original function is replaced by an N-Version implementation of that function.
The specific LLM and equivalence checker are external and configurable, which allows \toolname to use any off-the-shelf solutions.

\toolname generates a binary that contains an N-Version implementation of the reference function. This binary protects against miscompilation faults, thanks to  two properties:
(1) miscompilations are very sensitive to the input source code, e.g. they might be related to a specific keyword~\cite{regehr-volatile}; consequently, deriving program variants from diverse-but-equivalent source code is less likely to trigger miscompilations in every diversified instance, and;
(2) miscompilations are compiler-specific~\cite{yang2011finding}, meaning that it is improbable that variants produced by different compilers (see \autoref{sec:diversification}) share the same miscompilation error.

\subsection{Diversification Pass}
\label{sec:diversification}
The goal of this step is to systematically create function variants given a reference function.

As shown in \autoref{fig:pipeline-overview}, the Diversification pass is managed by a controller module.
The controller is responsible for reading configuration, interfacing with LLM APIs, and processing responses.
The invocation of the LLM is configured by the following parameters: the LLM API where the diversification request will be sent to; the reference function's source code; the number of variants to be generated; and a pair of input and output languages. The input language is the one of the reference function. The output language can be the same one as the reference function's (e.g. diversifying C-to-C), or a different language  (e.g. diversifying C-to-Go).

In \toolname, the LLM prompt structure is as follows:
\begin{itemize}
    \item A natural language description of the reference function: \textit{The following code is a reference implementation of a function in <INPUT>.}
    \item The reference function.
    \item A description of the task: \textit{Create <NUMBER> substitute implementation(s) of the function [in the <OUTPUT> language], which are different but equivalent. It should be possible to directly replace the function with any substitute, and it should provide the same functionality.}
    \item Additional remarks: \textit{Do not output any other text apart from code. Do not create auxiliary or helper functions. Maintain the original function's signature.}
\end{itemize}
The \textit{<NUMBER>}, \textit{<INPUT>}, and \textit{<OUTPUT>} strings are replaced according to the configuration, and the text in square brackets is omitted if the input and output languages are the same.
Note that some LLM APIs provide support for sampling n times with the same prompt (aka n\_sample). Our pilot experiment showed that a single prompt with n\_sample > 1 produced less diverse results than one single prompts requesting for  \textit{<NUMBER>} variant. Hence, this is what we use in this paper.

\begin{figure}[t]
    \centering
    \includegraphics[width=0.5\textwidth]{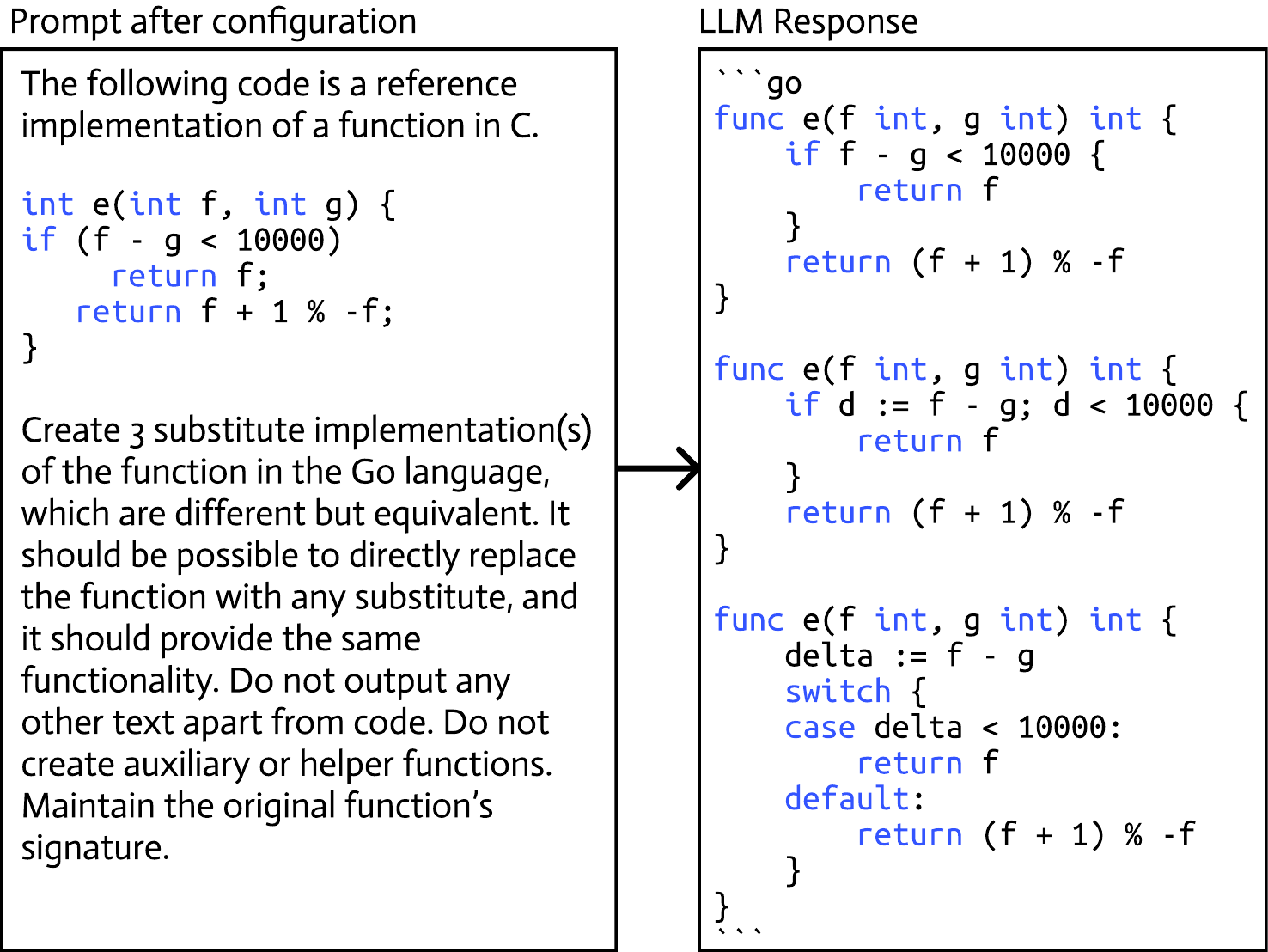}
    \caption{\toolname' diversification pass example. On the left box, the \toolname prompt; on the right box, the response from the LLM.}
    \label{fig:diversification}
\end{figure}

\autoref{fig:diversification} shows an example of the diversification pass, including the resulting prompt after configuration, and resulting function variants.
The configuration is set to generate three Go variants from a C reference function. Recall that translating to another language such as Go, provides runtime diversity which is the core goal of N-Version programming \cite{avizienis1985n}.

In the response, we can observe that while the first variant is a direct translation of the reference function, the second and third variants are dissimilar, using extra operations or different control flow constructs.

As shown in the previous example, the diversification pass can be performed in a cross-language configuration.
This is possible because the Validation and Harnessing passes of \toolname are performed at the IR level.
Hence, the source languages for the reference and the variants can be different, as long as there exist front ends that can compile the languages to the same IR.

This diversification approach presents two main challenges.
First, LLM outputs are not always correct~\cite{creative-correct}.
We need a mechanism to guarantee that the produced variants are equivalent.
This concern is addressed in the Validation pass (\autoref{sec:validation})
Second, when diversifying the reference to different programming languages, the IR code resulting from compiling the variants might not be compatible.
We mitigate this concern in the Harnessing pass (\autoref{sec:harnessing}).

To sum up, the Diversification pass of \toolname crafts an LLM prompt to produce  variants of the reference function either in the same language or in a different language that can be transformed into the same IR.
The resulting variants are forwarded to \toolname' next pass. 

\subsection{Validation Pass}
\label{sec:validation}

The goal of the Validation pass is to guarantee that the functionality of the function variants produced by the Diversification pass is equivalent to the reference.

\revised{In contrast to previous work where automatic diversification is achieved by performing transformations (1) that operate on low-level code representations, e.g. binary code, and (2) that are known to preserve equivalence~\cite{nop, kguard, wasm-mutate}, \toolname leverages LLMs at the source code-level, and thus the variants are not guaranteed to be correct by construction~\cite{NEURIPS2023_43e9d647, liu2024exploring}.}
To address this issue, \toolname performs four checks on each variant: compilation in isolation, compilation within containing project, test suite success, and equivalence check.

\textit{Compilation success in isolation:} In this check, \toolname inserts the variant into a file which only contains an entry point, e.g., a main function, that calls the variant with arbitrary parameters.
Then we attempt to compile this file with the corresponding language's toolchain.
The variant is passed to the next step if compilation completes without any error. 

\textit{Compilation success within the corresponding project:} In this check, \toolname replaces the reference function's IR inside the global project with the variant's IR.
The variant is filtered out if the build pipeline is unsuccessful.
Reasons for failure at this step are related to IR incompatibility, e.g. mismatching types.

\textit{Test suite success:} In this check, \toolname executes the test suite against each binary that includes a valid variant created in the previous step.
The variant is filtered out if any test from the suite does not pass.
At this step, a test case that fails is direct evidence of a functionality difference between the original and the variant.

\textit{Equivalence check:} In this final check, \toolname calls a configurable off-the-shelf formal equivalence checking tool, such as alive-tv~\cite{lopes2021alive2}, or Rust's Kani~\cite{kani}. 
This uncouples \toolname from any specific IR language and allows the use of any state-of-the-art solution for verification.
It performs the equivalence check with the reference and variant IR codes as parameters.
The variant is filtered out if the tool proves the variant to be non-equal, or if it fails to prove equivalence within a given time and memory budget.
Reasons for failure at this step are direct evidence of non-equivalence, as the variant deviates from the specification's functionality.
This is the most resource-consuming check, so it is executed last.
All function variants that pass the equivalence check are considered correct and can be forwarded to the next pass for assembly.
Addressing the accuracy of the equivalence checking tool is out of the scope of this work.
We point that equivalence check false negatives would not introduce errors to the system, as the variant would be filtered out.
Equivalence check false positives are more challenging, as non-equivalent variants would be accepted in the N-Version assembly, yet, their divergent output would not propagate silent corruptions, but instead would trigger a program termination (see ~\autoref{sec:harnessing}).

To sum up, the Validation pass filters candidate variants, eventually keeping only the variants for which it can provide formal correctness guarantees.
\toolname blends the radiating diversity powered by the LLM's creativity with strict guarantees of formal program equivalence checking.

\subsection{Harnessing Pass}
\label{sec:harnessing}
Per N-Version programming, the function variants need to be assembled into a single unit before execution.
In \toolname, the assembling is done at the intermediate representation level.

The previous passes of \toolname aim at maximizing diversity. They leverage the diversity of source languages and blend multiple toolchains.
Therefore, the variants' IR contain subtle differences, which is at the core of the software diversity we are seeking.
Yet, these differences need to be dealt with when assembling N-Version functions.
To handle this case, \toolname supports non-trivial IR transformations of the variants during the harnessing pass.
Specifically, \toolname implements Go-to-C transformations related to function signatures and replacing language-specific components: (1) the LLVM function's \texttt{nest} self-reference is removed; (2) the function's parameter names are normalized; and (3) IR calls to the Go "runtime" module are removed~\footnote{https://github.com/ASSERT-KTH/Galapagos/blob/main/linker/lib/go/go.cpp}.

\begin{figure}[t]
    \centering
    \includegraphics[width=0.65\textwidth]{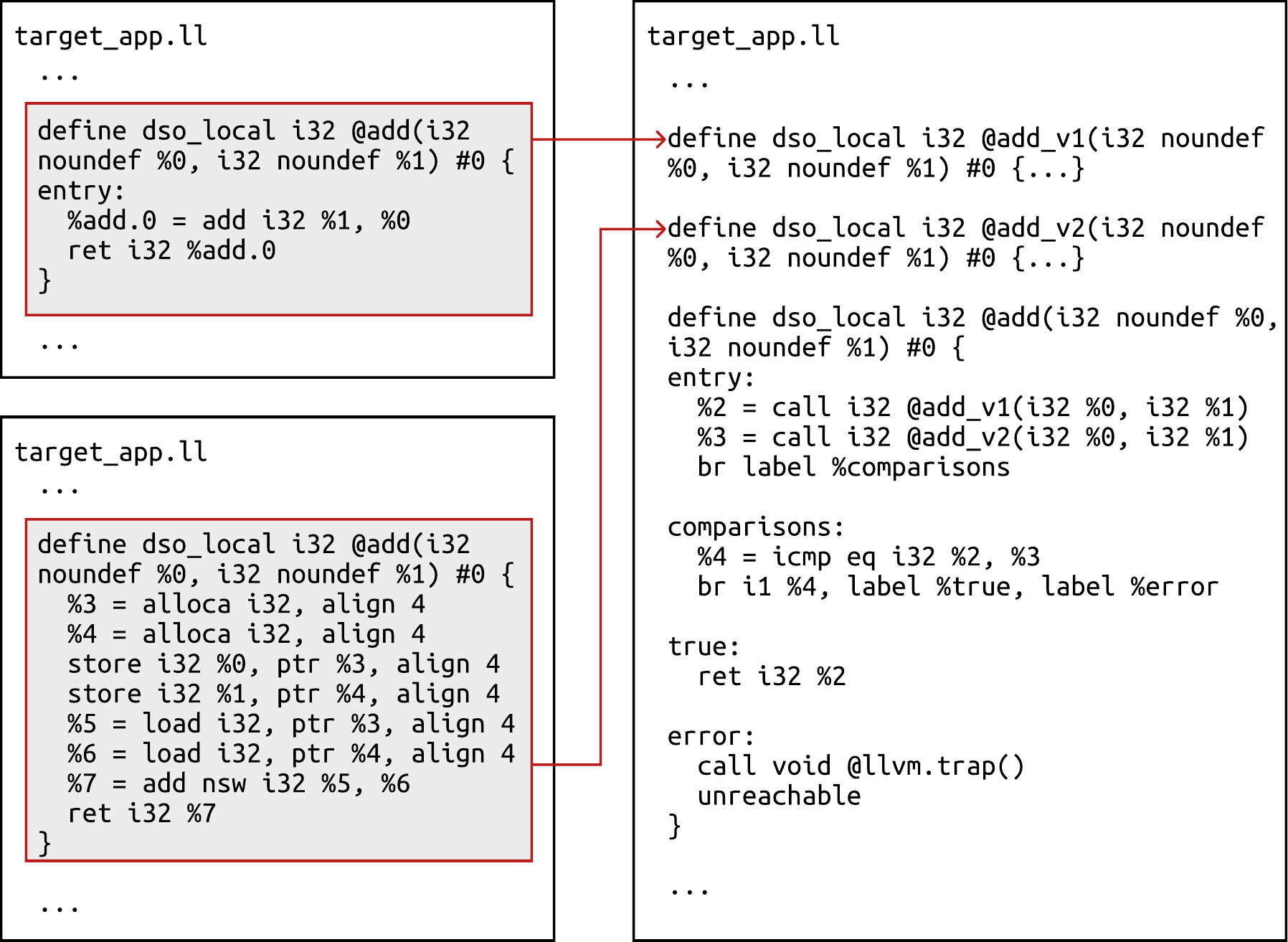}
    \caption{{\toolname}'s harnessing pass example, in LLVM IR.}
    \label{fig:harnessing-step}
\end{figure}

\autoref{fig:harnessing-step} shows an example where two equivalent variants of a function which adds two integers are harnessed into an N-Version function.
This example is written in the LLVM intermediate representation.
The variants' code is shown in the boxes on the left-hand side, and the resulting N-Version function is shown in the box on the right-hand side.
In the N-Version form, the code from the single variants is inserted in the corresponding file as a function definition.
These definitions are named after the original function plus a prefix.
\toolname then generates a wrapper function named after the original reference function. This function invokes all the different variants and compares their outputs.
\toolname also transforms the original application to redirect all existing invocations of the reference to the wrapper function.
Finally, the binary is produced from the resulting IR code.

For N-Version programs, the output from the N functions can be selected in different ways. In \toolname, we support N-of-N selection~\cite{hydra}, i.e. we compare the return values of each variant against each other and return a value only if all the variants agree, otherwise \toolname forcefully terminates the execution.

\subsection{Class of  Functions Considered for Automated N-Version}
\label{sec:functions}
\toolname transforms functions into N-Version functions, with equivalence guarantees. To achieve this challenging task, we must assume a specific class of functions.
Namely, these functions must be pure, i.e. must deterministically produce the same output from the same input, and must not cause any side-effect.
This is a requisite to our N-Version approach since all function versions must be executed simultaneously, and their outputs compared.
Only with pure functions can we ensure that multiple executions of the function will not result in the accumulation of undesired side effects.

This scope for \toolname is relevant, as pure functions represent a significant subset of all functions in software~\cite{pure-java, pure-javascript}.
\autoref{fig:function-class} shows a relevant example from the cryptography library \href{https://github.com/jedisct1/libsodium/blob/6e27e98777f3a6eeb233cd3280412414cead4e0d/src/libsodium/crypto_scalarmult/ed25519/ref10/scalarmult_ed25519_ref10.c#L9}{libsodium} that can be divsersified.
Furthermore, we argue that in practice, critical sections of code that have strong reliability requirements can be refactored into pure functions, making hardening via \toolname and N-Version programming possible.

\begin{figure}
    \centering
    \begin{lstlisting}[style=Cstyle]
static int
_crypto_scalarmult_ed25519_is_inf(const unsigned char s[32])
{
    unsigned char c;
    unsigned int  i;

    c = s[0] ^ 0x01;
    for (i = 1; i < 31; i++) {
        c |= s[i];
    }
    c |= s[31] & 0x7f;

    return ((((unsigned int) c) - 1U) >> 8) & 1;
}        
    \end{lstlisting}
    \caption{Pure function example from the cryptography library libsodium, which meets the criteria for automated N-Version programming with \toolname.}
    \label{fig:function-class}
\end{figure}

\subsection{Implementation}

We implement the Diversification and Validation passes of \toolname in Python.

The LLM used in our experiments is OpenAI's GPT-4o, because it was the best at the time of doing the experiments of this paper. The pipeline would support plugging other LLMs. It is left to future work to study the impact of varying the LLM on the diversification pipeline.

For formal equivalence checks, we rely on the \texttt{alive-tv}~\cite{lopes2021alive2} tool that checks equivalence at the LLVM IR level.
\texttt{alive-tv} is an equivalence checker tied to the LLVM IR and its semantics.
It operates by comparing two versions of a function: source and target.
These versions must be well-formed LLVM IR and adhere to the same LLVM specification version.
It verifies that the target function preserves the semantics of the source function, and enforces that the target function must be a refinement of the source function.
This is, for all inputs where the source function has defined behavior, the target must produce the same observable result.
The comparison is made by translating both variants to SMT and asks for logical equivalence. If the problem is satisfiable, it means that the solver has found a counter-example, that is an input such that both variants behave differently.

The Harnessing pass is implemented in C++ in order to make use of LLVM's code manipulation libraries. It programmatically combines the LLVM IR version of the variants and calls the N-Version output comparison function.

\toolname supports C as input language, and C and Go as output languages.
Support can be extended to any language pair, as long as these can be compiled to a common intermediate representation. 

To provide maintanability and debugging in \toolname, all outputs from each pass are saved to disk.
This includes raw data from the LLM call, all variants' source code and all resulting IR and machine code. 

The implementation of \toolname can be found in our open source repository at \url{https://github.com/ASSERT-KTH/Galapagos}.

\section{Experimental Methodology}
\label{sec:methodology}

In this section, we introduce the research questions that structure our empirical validation of \toolname. 

\subsection{Research Questions}

To systematically evaluate the effectiveness of \toolname, we define the following research questions: 

\vspace{7pt}

\begin{enumerate}
    \item[\textbf{RQ1}] \textit{To what extent is a large language model able to produce functionally equivalent variants of programs at the source-code level?}
    
    \vspace{7pt}
    
    The variants generated by an LLM come with no guarantee. We aim to compute what percentage of these variants can be proven to perform the same task as the original function under diversification.
    To answer this question, we take a representative sample of functions from relevant C applications and libraries, and run them through \toolname' pipeline.
    Then, we analyze the distribution of these variants in terms of the different correctness filters of \toolname' validation passes presented in \autoref{sec:validation}.

    \vspace{7pt}
    
    \item[\textbf{RQ2}] \textit{To what extent do the program variants produced by a large language model exhibit diverse observable internal behaviors?}
    
    \vspace{7pt}
    
    The variants generated by an LLM can be different at the source code or IR code.
    However, program variants stored on disk do not increase reliability per se. 
    N-Version design increases reliability when variants exhibit observable differences. 
    For this, we compute how much of the variants' diversity is preserved after all compilation and optimization stages.
    Furthermore, we compare how much CPU instructions vary between different function variants.
    \\
    \item[\textbf{RQ3}] \textit{To what extent are equivalent variants produced by a large language model useful to harden programs against the considered fault model?}
    
    \vspace{7pt} 
    We evaluate the effectiveness of N-Version functions per our fault model, miscompilation bugs (see \autoref{sec:fault-model}). We curate publicly documented, reproducible miscompilation bugs, and determine whether LLM-generated variants are useful in detecting and preventing them.
    \\
    
    \item[\textbf{RQ4}] \textit{What is the performance overhead introduced by the N-Version functions generated by \toolname?}

    \vspace{7pt}
    N-Version functions with high execution overhead might be impractical.
    Therefore, we measure the resulting performance overhead introduced by \toolname. 
    To do this, we run the original and N-Version functions and compare their average execution times.

\end{enumerate}

\subsection{Dataset of Functions}
\label{sec:dataset}
We select C functions from six open-source projects: alsa-lib, FFmpeg, libgcrypt, liboqs, libsodium, openssl. Those projects are all real-world, and have differing sizes, stages of maturity, and adoption.
~\autoref{tab:projects} shows the selected projects, their versions, and related statistics. 
The number of commits for the selected projects' repositories ranges between \numprint{1458} and \numprint{111662}.
Likewise, the number of C files and declared functions goes from \numprint{349} to \numprint{4160}, and \numprint{3683} to \numprint{110532} respectively. 
\begin{table}[]
    \centering
    \rowcolors{2}{gray!10}{white}
    \begin{tabular}{lcrrr}
        \toprule
        \textbf{Project} & \textbf{Version} & \textbf{\# Commits} & \textbf{\# Files}  & \textbf{\# Functions} \\
        \midrule
        \href{https://github.com/alsa-project/alsa-lib/tree/3864f7d95f7ad97690ed8e823aab3c96abfe5e48}{alsa-lib}      & v1.2.11* & \numprint{4445} & \numprint{400} & \numprint{4004} \\
        \href{https://github.com/FFmpeg/FFmpeg/tree/567e78b283}{FFmpeg}      & n6.1-dev & \numprint{111662} & \numprint{4160} & \numprint{110532} \\
        \href{https://github.com/gpg/libgcrypt/tree/ad3b599462bdbc459f6c7be867e9a12ab46481b3}{libgcrypt}   & 1.10-base* & \numprint{3419}   & \numprint{588} & \numprint{5616} \\
        \href{https://github.com/open-quantum-safe/liboqs/tree/a5ec23cf19763d36a558b8358345823ae45d57e5}{liboqs}   & 0.10.0* & \numprint{1458}   & \numprint{3625} & \numprint{9594} \\
        \href{https://github.com/jedisct1/libsodium/tree/d2ac311e}{libsodium}   & 1.0.18 & \numprint{4204}   & \numprint{349} & \numprint{3683} \\
        \href{https://github.com/openssl/openssl/tree/42a6a25ba4}{openssl}     & 3.0.0-beta2 & \numprint{32959}  & \numprint{4004} & \numprint{94984} \\
        \bottomrule
    \end{tabular}
    \caption{List of real-world C projects used for experimental validation,. The versions marked with (*) represent the oldest named version before the commit used in this dataset.}
    \label{tab:projects}
\end{table}

From each of these projects, we select the first five functions that meet the following properties, analyzed sequentially.
(1) It is a pure function, and;
(2) It does not call other functions.
These properties ensure that the functions are supported by the semantic equivalence tool used in \toolname' verification step.

\begin{table}[]
    \centering
    \rowcolors{2}{gray!10}{white}
    \begin{tabular}{clrr}
        \toprule
        \textbf{Project} & \textbf{Function Name} & \textbf{Size~(LOC)} & \textbf{\# Refs.} \\
        \midrule
          \cellcolor{white} & \href{https://github.com/alsa-project/alsa-lib/blob/3864f7d95f7ad97690ed8e823aab3c96abfe5e48/src/pcm/pcm_alaw.c#L125}{alaw\_to\_s16}   & \numprint{15} & \numprint{1}\\
          \cellcolor{white} & \href{https://github.com/alsa-project/alsa-lib/blob/3864f7d95f7ad97690ed8e823aab3c96abfe5e48/src/pcm/pcm_iec958.c#L77}{iec958\_parity}   & \numprint{13} & \numprint{1}\\
          \cellcolor{white} & \href{https://github.com/alsa-project/alsa-lib/blob/3864f7d95f7ad97690ed8e823aab3c96abfe5e48/src/pcm/interval.c#L74}{add}   & \numprint{6} & \numprint{3}\\
          \cellcolor{white} & \href{https://github.com/alsa-project/alsa-lib/blob/3864f7d95f7ad97690ed8e823aab3c96abfe5e48/src/pcm/pcm_mulaw.c#L141}{ulaw\_to\_s16}   & \numprint{15} & \numprint{1}\\
         
        \multirow{-5}{*}{alsa-lib} \cellcolor{white} &   \href{https://github.com/alsa-project/alsa-lib/blob/3864f7d95f7ad97690ed8e823aab3c96abfe5e48/src/pcm/pcm_alaw.c#L59}{val\_seg} & \numprint{15} & \numprint{2}\\
        \midrule
          \cellcolor{white} &   \href{https://github.com/FFmpeg/FFmpeg/blob/567e78b28320939c18e16acbdbeb2b77d24e2c03/libavcodec/flacenc.c#L166}{flac\_get\_max\_frame\_size} & \numprint{19} & \numprint{2}\\
          \cellcolor{white} &   \href{https://github.com/FFmpeg/FFmpeg/blob/567e78b28320939c18e16acbdbeb2b77d24e2c03/libavcodec/4xm.c#L717}{mix} & \numprint{6} & \numprint{43}\\
          \cellcolor{white} &   \href{https://github.com/FFmpeg/FFmpeg/blob/567e78b28320939c18e16acbdbeb2b77d24e2c03/libavcodec/vp56dsp.c#L48}{vp5\_adjust*} & \numprint{14} & \numprint{2}\\
          \cellcolor{white} &   \href{https://github.com/FFmpeg/FFmpeg/blob/567e78b28320939c18e16acbdbeb2b77d24e2c03/libavcodec/diracdec.c#L1562}{weight} \cellcolor{white} & \numprint{10} & \numprint{2}\\
         
        \multirow{-5}{*}{FFmpeg} \cellcolor{white} &   \href{https://github.com/FFmpeg/FFmpeg/blob/567e78b28320939c18e16acbdbeb2b77d24e2c03/libavfilter/vf_rotate.c#L202}{int\_sin} & \numprint{17} & \numprint{5}\\
        \midrule
          \cellcolor{white} &   \href{https://github.com/gpg/libgcrypt/blob/ad3b599462bdbc459f6c7be867e9a12ab46481b3/cipher/kyber-common.c#L759}{barrett\_reduce} & \numprint{7} & \numprint{2}\\
          \cellcolor{white} &   \href{https://github.com/gpg/libgcrypt/blob/ad3b599462bdbc459f6c7be867e9a12ab46481b3/cipher/mceliece6688128f.c#L3228}{ctz} & \numprint{12} & \numprint{1}\\
          \cellcolor{white} &   \href{https://github.com/gpg/libgcrypt/blob/ad3b599462bdbc459f6c7be867e9a12ab46481b3/cipher/sntrup761.c#L386}{int16\_t\_negative\_mask} & \numprint{7} & \numprint{2}\\
          \cellcolor{white} &   \href{https://github.com/gpg/libgcrypt/blob/ad3b599462bdbc459f6c7be867e9a12ab46481b3/cipher/sntrup761.c#L375}{int16\_t\_nonzero\_mask} & \numprint{7} & \numprint{5}\\
         
        \multirow{-5}{*}{libgcrypt} \cellcolor{white} &   \href{https://github.com/gpg/libgcrypt/blob/ad3b599462bdbc459f6c7be867e9a12ab46481b3/cipher/kyber-common.c#L740}{montgomery\_reduce} & \numprint{7} & \numprint{2}\\
        \midrule
          \cellcolor{white} & \href{https://github.com/open-quantum-safe/liboqs/blob/a5ec23cf19763d36a558b8358345823ae45d57e5/src/sig/falcon/pqclean_falcon-padded-1024_clean/fpr.h#L403}{fpr\_half} & \numprint{11} & \numprint{132}\\
          \cellcolor{white} & \href{https://github.com/open-quantum-safe/liboqs/blob/a5ec23cf19763d36a558b8358345823ae45d57e5/src/sig/falcon/pqclean_falcon-padded-1024_clean/fpr.h#L447}{fpr\_lt} & \numprint{29} & \numprint{80}\\
          \cellcolor{white} & \href{https://github.com/open-quantum-safe/liboqs/blob/a5ec23cf19763d36a558b8358345823ae45d57e5/src/sig/falcon/pqclean_falcon-1024_avx2/keygen.c#L716}{modp\_montymul} & \numprint{9} & \numprint{1220}\\
          \cellcolor{white} & \href{https://github.com/open-quantum-safe/liboqs/blob/a5ec23cf19763d36a558b8358345823ae45d57e5/src/sig/falcon/pqclean_falcon-1024_avx2/keygen.c#L643}{modp\_norm} & \numprint{2} & \numprint{80}\\
         
        \multirow{-5}{*}{liboqs} \cellcolor{white} & \href{https://github.com/open-quantum-safe/liboqs/blob/a5ec23cf19763d36a558b8358345823ae45d57e5/src/kem/ntruprime/pqclean_sntrup761_clean/crypto_core_wforcesntrup761.c#L14}{int16\_nonzero\_mask} & \numprint{6} & \numprint{16}\\
        \midrule
          \cellcolor{white} & \href{https://github.com/jedisct1/libsodium/blob/d2ac311e0eace877fda1a59a573f26c0b6b70435/src/libsodium/sodium/codecs.c#L118}{b64\_byte\_to\_char} & \numprint{6} & \numprint{4}\\
          \cellcolor{white} & \href{https://github.com/jedisct1/libsodium/blob/d2ac311e0eace877fda1a59a573f26c0b6b70435/src/libsodium/sodium/codecs.c#L139}{b64\_byte\_to\_urlsafe\_char} & \numprint{6} & \numprint{4}\\
          \cellcolor{white} & \href{https://github.com/jedisct1/libsodium/blob/d2ac311e0eace877fda1a59a573f26c0b6b70435/src/libsodium/sodium/codecs.c#L127}{b64\_char\_to\_byte} & \numprint{9} & \numprint{2}\\
          \cellcolor{white} & \href{https://github.com/jedisct1/libsodium/blob/d2ac311e0eace877fda1a59a573f26c0b6b70435/src/libsodium/sodium/codecs.c#L148}{b64\_urlsafe\_char\_to\_byte} & \numprint{9} & \numprint{2}\\
         
        \multirow{-5}{*}{libsodium} \cellcolor{white} &   \href{https://github.com/jedisct1/libsodium/blob/d2ac311e0eace877fda1a59a573f26c0b6b70435/src/libsodium/crypto_pwhash/argon2/blamka-round-ref.h#L8}{fBlaMka} & \numprint{5} & \numprint{28}\\
        \midrule
          \cellcolor{white} & \href{https://github.com/openssl/openssl/blob/42a6a25ba4ddb40333e92e6e2fc57625d9567090/crypto/rsa/rsa_lib.c#L244}{icbrt64} & \numprint{15} & \numprint{1}\\
          \cellcolor{white} & \href{https://github.com/openssl/openssl/blob/42a6a25ba4ddb40333e92e6e2fc57625d9567090/crypto/sha/keccak1600.c#L1021}{BitDeinterleave} & \numprint{34} & \numprint{1}\\
          \cellcolor{white} & \href{https://github.com/openssl/openssl/blob/42a6a25ba4ddb40333e92e6e2fc57625d9567090/crypto/sha/keccak1600.c#L985}{BitInterleave} & \numprint{34} & \numprint{1}\\
         \cellcolor{white} & \href{https://github.com/openssl/openssl/blob/42a6a25ba4ddb40333e92e6e2fc57625d9567090/crypto/ec/ecp_nistz256.c#L153}{\_booth\_recode\_w5} & \numprint{10} & \numprint{3}\\
        \multirow{-5}{*}{openssl} \cellcolor{white} & \href{https://github.com/openssl/openssl/blob/42a6a25ba4ddb40333e92e6e2fc57625d9567090/crypto/ec/ecp_nistz256.c#L165}{\_booth\_recode\_w7} & \numprint{10} & \numprint{2}\\
        \bottomrule
    \end{tabular}
    \caption{List of 30 candidate real-world C functions for diversification. LOC "stands for lines of code". \# Refs. stands for number of times that the function is statically referenced within the project.}
    \label{tab:functions}
\end{table}

\autoref{tab:functions} describes the set of functions we selected, which covers a diverse range of functionalities, e.g. arithmetic, type conversion, and cryptographic operations.
It also shows each function's size in lines of code (LOC), and the number of times that it is statically referenced within the project (\#Refs).
The size of the functions by lines of code ranges from \numprint{2} to \numprint{34} with a median of \numprint{10}.
The number of times that each function is called within the same project ranges from \numprint{1} to \numprint{1220} with a median of \numprint{2}.
To sum up, this dataset is exclusively composed of real-world C code that is used daily by millions of users.

\begin{table*}[]
    \centering
    \rowcolors{2}{gray!10}{white}
    \begin{tabular}{lllll}
    \toprule
    Tag & Version & Description & Issue \# & Status \\
    \midrule
    \href{https://github.com/ASSERT-KTH/Galapagos/blob/main/miscompilation/bug-1/project/main.c}{M1} & 15.0.0 & Wrong code with \texttt{-O1}& \href{https://github.com/llvm/llvm-project/issues/58765}{58765} & Unresolved \\
    \href{https://github.com/ASSERT-KTH/Galapagos/blob/main/miscompilation/bug-2/project/main.c}{M2} & 17.0.0 & Wrong code with \texttt{-Os} & \href{https://github.com/llvm/llvm-project/issues/68871}{68871} & Unresolved \\
    \href{https://github.com/ASSERT-KTH/Galapagos/blob/main/miscompilation/bug-3/project/main.c}{M3} & 18.0.0 & Wrong code with \texttt{-march=znver4 -flto -O3} & \href{https://github.com/llvm/llvm-project/issues/80494}{80494} & Unresolved \\ 
    \bottomrule
         
    \end{tabular}
    \caption{List of known miscompilation bugs selected for mitigation through \toolname.}
    \label{fig:miscompilations}
\end{table*}

\subsection{Methodology for RQ1: Assessing Functions Diversified by LLMs}
\label{sec:rq1-methodology}

As part of its Diversification pass, \toolname generates an arbitrary number of raw variants.
However, there are mandatory properties which every variant needs to comply with for N-Version design:
compilability, test suite execution success, and formal equivalence assessment.
To have a detailed view of the distribution of equivalent and non-equivalent variants, we measure the proportion of variants which fail at each filter and the reasons behind the failures.
Ultimately, the effectiveness of LLM-based diversification is measured by the proportion of variants which pass all filters and are proven equivalent.

Importantly, we diversify the dataset of C functions~(\autoref{sec:dataset}) with two configurations same-language (C-to-C) and cross-language (C-to-Go). 
We produce 10 variants for each configuration, i.e. 100 variants per project, and 600 variants in total.
Each variant is checked against all the filters described above, and the pass/fail occurrences are reported.

\subsection{Methodology for RQ2: Assessing Variants Uniqueness}
\label{sec:rq2-methodology}

To answer RQ2, we assess the diversity of the variants both statically and dynamically. 

Diversity at source code level can be removed through the transformations performed at different stages of compilation~\cite{cabrera-thesis}. We measure how often this occurs, collecting the proportion of variants that preserve uniqueness after compilation with different optimization flags.
We compare the unique SHA-256 hashes of the produced machine code, assuming that compilation is repeatable.
This experiment is performed on the same dataset of C functions as in RQ1~(\ref{sec:dataset}).
We perform this comparison only for variants that have been proven functionally equivalent previously by the filters applied in \ref{sec:rq1-methodology}.
We use each of four optimization flags \texttt{-O0}, \texttt{-O1}, \texttt{-O2}, \texttt{-O3} to rebuild the function's machine code, and compare the SHA256 hash of the produced binaries, to inspect whether some compilation flags undo specific variants (i.e. make the binary statically identical to the original).

Then, we execute all statically diverse variants, to measure the diversity of observable behaviors among them. 
We define diverse observable behaviors per the CPU instruction trace.
At runtime, we record all CPU instructions executed by the variant.
To achieve this, we rely on Intel's Pin instrumentation tool~\cite{intel-pin}, since it allows us to record CPU instructions at function granularity, filtering out all other execution inherent to the used runtime, e.g. state initialization, garbage collection, etc.
We then aggregate and compare the instructions executed by each variant in two ways.
First, we compare the CPU instruction set used by the variants and compare it against the set of the original CPU instruction set.
This is measured using the Jaccard similarity coefficient~\cite{jaccard}.
Second, we compare the amount of CPU instructions executed by the variants in terms of absolute numbers.

\revised{Finally, we compare the dynamic diversity of the LLM-generated variants against a baseline of rule-based variants generated with state-of-the-art tool Tigress~\cite{tigress}. Tigress is a code transformation tool developed and maintained by Collberg and colleagues \cite{collberg2012distributed} for diversification and obfuscation.
We consider Tigress a suitable baseline approach since: (1) it works at the same level of abstraction; and (2) it aims to achieve behavioral diversity through program transformation, aligning with the objectives of N-Version programming. 
We apply the control-flow flattening transformation and use the resulting variants for comparison at runtime using the Jaccard similarity coefficient.}

\subsection{Methodology for RQ3: Validating the Ability to Mitigate Miscompilations}
\label{sec:rq3-methodology}
As described in~\autoref{sec:fault-model}, our work specifically focuses on hardening against miscompilation errors produced by bugs in compilers, which is a very dangerous and tricky class of bugs.
In RQ3, we consider miscompilation bugs from the Clang compiler.
We assess the effectiveness of \toolname in automatically generating N-Version programs which can mitigate these bugs.
We look at all bugs reported in \autoref{fig:miscompilations}, coming from different versions of Clang and optimization stages. All are unresolved in the repository's issue tracker at the time of experimentation.
For each bug, we create two test programs:
one with a function containing a minimal reproducible example for said bug, and one with an N-Version implementation of the same function.
We always run \toolname with the minimal function as the reference broken run.
We consider \toolname to be successful if the N-Version implementation of the reference forces a crash instead of producing the incorrect output due to the miscompilation bug.

\subsection{Methodology for RQ4: Measuring Performance Overhead}
\label{sec:rq4-methodology}
To measure the performance overhead of N-Version functions generated by \toolname, we use the \texttt{clock\_gettime(3)} API.
To get an accurate comparison, we time both the original single-version and the N-Version functions over \numprint{2000000} executions. 
Execution and measurements are done in an Intel i7-11800H system, with 16 GB of RAM.

\section{Experimental Results}
\label{sec:results}
In this section, we present and discuss the results of our experiments. In \autoref{sec:res-validation} we look at the ability of \toolname to generate variants that are equivalent to the reference (RQ1). In \autoref{sec:static-uniqueness} and \autoref{sec:dynamic-uniqueness}, we assess the static and dynamic diversity of the variants, to answer RQ2. We conclude with \autoref{sec:mitigation}, where we answer RQ3 with empirical evidence about the resulting N-Version units' ability to mitigate miscompilation errors.

\subsection{Validating Variants}
\label{sec:res-validation}

\begin{table*}[]
    \centering
    \rowcolors{5}{white}{gray!10}
    \begin{tabular}{llrrrr|rrrr}
        \toprule
        \multicolumn{1}{c}{\multirow{2}{*}{\textbf{Project}}} & \multicolumn{1}{c}{\multirow{2}{*}{\textbf{Function Name}}} & \multicolumn{4}{c}{\textbf{ \makecell{Same-language \\ diversification (C)}}} & \multicolumn{4}{c}{\textbf{\makecell{Cross-language \\ diversification (C to Go)}}}  \\
\cmidrule(lr){3-6} \cmidrule(lr){7-10}
        {} & {} & \footnotesize \textbf{\makecell{Comp. \\ Isol.}} & \footnotesize \textbf{\makecell{Comp.\\ Proj.}} & \footnotesize \textbf{Test} & \footnotesize \textbf{Verif.} & \footnotesize \textbf{\makecell{Comp. \\ Isol.}} & \footnotesize \textbf{\makecell{Comp.\\ Proj.}} & \footnotesize \textbf{Test.} & \footnotesize \textbf{Verif.} \\
        \midrule
\cellcolor{white} & alaw\_to\_s16 & \numprint{10} & \numprint{10} & \numprint{10} & \numprint{1} & \numprint{10} & \numprint{10} & \numprint{10} & \numprint{6}  \\
\cellcolor{white} & iec958\_parity & \numprint{10} & \numprint{8} & \numprint{8} & \numprint{8} & \numprint{10} & \numprint{10} & \numprint{10} & \numprint{10}  \\
\cellcolor{white} & add & \numprint{10} & \numprint{9} & \numprint{9} & \numprint{9} & \numprint{10} & \numprint{1} & \numprint{1} & \numprint{1}  \\
\cellcolor{white} & ulaw\_to\_s16 & \numprint{10} & \numprint{10} & \numprint{10} & \numprint{9} & \numprint{10} & \numprint{10} & \numprint{10} & \numprint{10}  \\
\cellcolor{white}
\multirow{-5}{*}{alsa-lib} & val\_seg & \numprint{10} & \numprint{10} & \numprint{10} & \numprint{2} & \numprint{9} & \numprint{1} & \numprint{1} & \numprint{0}  \\
\midrule
\cellcolor{white} & flac\_get\_max\_frame\_size & \numprint{10} & \numprint{10} & \numprint{9} & \numprint{4} & \numprint{10} & \numprint{10} & \numprint{10} & \numprint{0}  \\
\cellcolor{white} & mix & \numprint{10} & \numprint{10} & \numprint{10} & \numprint{6} & \numprint{10} & \numprint{10} & \numprint{10} & \numprint{0}  \\
\cellcolor{white} & vp5\_adjust & \numprint{10} & \numprint{10} & \numprint{10} & \numprint{0} & \numprint{6} & \numprint{0} & \numprint{0} & \numprint{0}  \\
\cellcolor{white} & weight & \numprint{10} & \numprint{10} & \numprint{10} & \numprint{10} & \numprint{10} & \numprint{10} & \numprint{10} & \numprint{0}  \\
\cellcolor{white}
\multirow{-4}{*}{ffmpeg} & int\_sin & \numprint{10} & \numprint{10} & \numprint{10} & \numprint{0} & \numprint{10} & \numprint{10} & \numprint{10} & \numprint{0}  \\
\midrule
\cellcolor{white} & barrett\_reduce & \numprint{10} & \numprint{10} & \numprint{0} & \numprint{0} & \numprint{10} & \numprint{10} & \numprint{10} & \numprint{10}  \\
\cellcolor{white} & ctz & \numprint{10} & \numprint{10} & \numprint{10} & \numprint{9} & \numprint{10} & \numprint{10} & \numprint{10} & \numprint{6}  \\
\cellcolor{white} & int16\_t\_negative\_mask & \numprint{9} & \numprint{9} & \numprint{7} & \numprint{7} & \numprint{10} & \numprint{8} & \numprint{6} & \numprint{5}  \\
\cellcolor{white} & int16\_t\_nonzero\_mask & \numprint{10} & \numprint{10} & \numprint{7} & \numprint{7} & \numprint{9} & \numprint{9} & \numprint{5} & \numprint{5}  \\
\cellcolor{white}
\multirow{-5}{*}{libgcrypt} & montgomery\_reduce & \numprint{10} & \numprint{10} & \numprint{6} & \numprint{4} & \numprint{10} & \numprint{10} & \numprint{8} & \numprint{8}  \\
\midrule
\cellcolor{white} & fpr\_half & \numprint{10} & \numprint{10} & \numprint{10} & \numprint{2} & \numprint{10} & \numprint{10} & \numprint{10} & \numprint{1}  \\
\cellcolor{white} & fpr\_lt & \numprint{9} & \numprint{9} & \numprint{9} & \numprint{0} & \numprint{9} & \numprint{9} & \numprint{9} & \numprint{1}  \\
\cellcolor{white} & modp\_montymul & \numprint{10} & \numprint{10} & \numprint{10} & \numprint{2} & \numprint{10} & \numprint{10} & \numprint{9} & \numprint{0}  \\
\cellcolor{white} & modp\_norm & \numprint{10} & \numprint{10} & \numprint{7} & \numprint{0} & \numprint{8} & \numprint{8} & \numprint{3} & \numprint{2}  \\
\cellcolor{white}
\multirow{-5}{*}{liboqs} & int16\_nonzero\_mask & \numprint{10} & \numprint{10} & \numprint{10} & \numprint{3} & \numprint{10} & \numprint{10} & \numprint{10} & \numprint{8}  \\
\midrule
\cellcolor{white} & b64\_byte\_to\_char & \numprint{10} & \numprint{8} & \numprint{8} & \numprint{0} & \numprint{10} & \numprint{6} & \numprint{6} & \numprint{0}  \\
\cellcolor{white} & b64\_byte\_to\_urlsafe\_char & \numprint{10} & \numprint{8} & \numprint{8} & \numprint{0} & \numprint{10} & \numprint{9} & \numprint{9} & \numprint{0}  \\
\cellcolor{white} & b64\_char\_to\_byte & \numprint{10} & \numprint{7} & \numprint{0} & \numprint{0} & \numprint{9} & \numprint{8} & \numprint{0} & \numprint{0}  \\
\cellcolor{white} & b64\_urlsafe\_char\_to\_byte & \numprint{10} & \numprint{10} & \numprint{0} & \numprint{0} & \numprint{8} & \numprint{0} & \numprint{0} & \numprint{0}  \\
\cellcolor{white}
\multirow{-5}{*}{libsodium} & fBlaMka & \numprint{10} & \numprint{10} & \numprint{10} & \numprint{9} & \numprint{10} & \numprint{10} & \numprint{10} & \numprint{0}  \\
\midrule
\cellcolor{white} & icbrt64 & \numprint{10} & \numprint{10} & \numprint{10} & \numprint{7} & \numprint{9} & \numprint{9} & \numprint{9} & \numprint{0}  \\
\cellcolor{white} & BitDeinterleave & \numprint{10} & \numprint{10} & \numprint{10} & \numprint{0} & \numprint{10} & \numprint{10} & \numprint{10} & \numprint{3}  \\
\cellcolor{white} & BitInterleave & \numprint{10} & \numprint{10} & \numprint{10} & \revised{\numprint{1}} & \numprint{1} & \numprint{1} & \numprint{1} & \numprint{1}  \\
\cellcolor{white} & \_booth\_recode\_w5 & \numprint{10} & \numprint{10} & \numprint{10} & \numprint{2} & \numprint{10} & \numprint{10} & \numprint{10} & \numprint{2}  \\
\cellcolor{white}
\multirow{-5}{*}{openssl} & \_booth\_recode\_w7 & \numprint{10} & \numprint{10} & \numprint{10} & \numprint{0} & \numprint{10} & \numprint{0} & \numprint{0} & \numprint{0}  \\
\midrule
        \multicolumn{1}{c}{avg.} & {} & 99.3\% & 96.0\% & 82.6\% & \revised{34.0\%} & 92.6\% & 76.3\% & 69.0\% & 24.3\% \\
        
        \bottomrule
    \end{tabular}
    \caption{Count of variant compliance to each validation filter: Compiled in isolation (Comp. Isol.), compiled within project (Comp. Proj.), test suite success (Test), and verified equal (Verif.).}
    \label{tab:results-rq1}
\end{table*}

To answer RQ1, we analyze the extent to which the generated variants pass the different validation filters. The numbers are presented in \autoref{tab:results-rq1}.

\begin{figure}[t]
    \centering
    \includegraphics[width=0.4\textwidth]{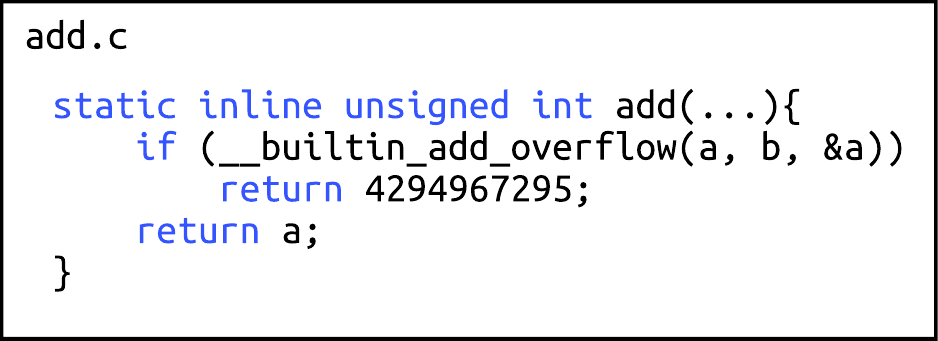}
    \caption{Build failure example: Function \texttt{`\_\_builtin\_add\_overflow}` is not declared in the original project's IR.}
    \label{fig:within-fail}
\end{figure}

First, we discuss the results for the generation of variants, in the same language. 
On average, \numprint{99.3}\% (298/300) of the variants are compiled correctly in isolation, and, when inserted in the project, \numprint{96.0}\% (288/300) still successfully compile.
Compilation in isolation fails in 2 out of 300 cases: one case because of a syntax error, and the second because the variant calls a function from a library that has not been imported.
Compilation within the project fails in 10 out of 298 remaining cases, all of them because the variant's IR code makes use of a function that is not declared in the project's IR code.
For instance, one of the variants of alsa-lib's \texttt{add} uses a compiler built-in function, and compiles in isolation, shown in \autoref{fig:within-fail}.
Yet, it fails to compile inside the project, as the built-in function is not declared in the IR scope of the reference function.

\begin{figure}[t]
    \centering
    \includegraphics[width=0.4\textwidth]{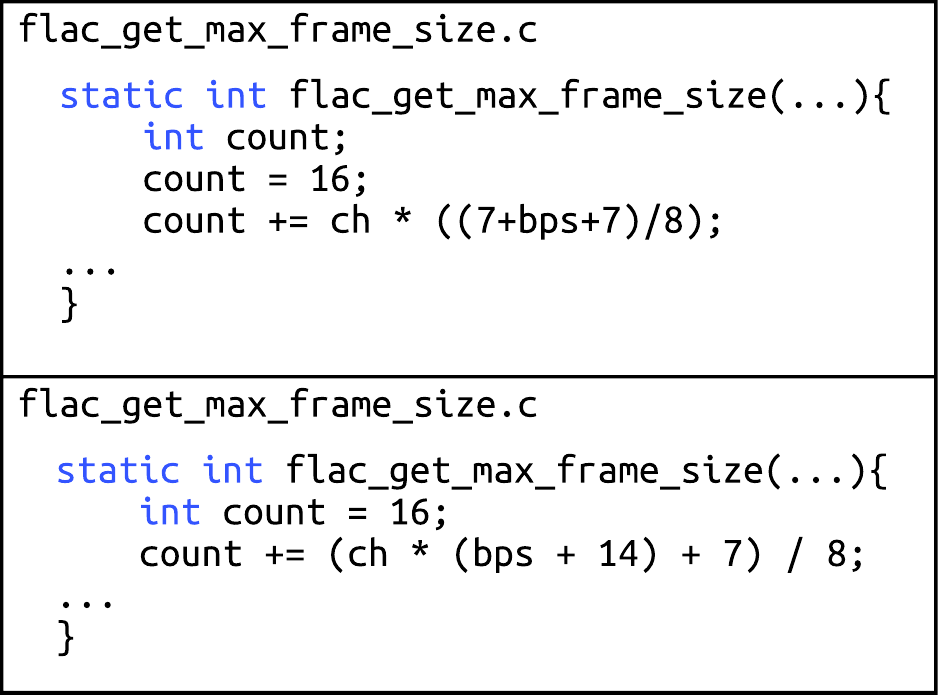}
    \caption{A logical error in variant (bottom) which causes test failure. The addition to \texttt{count} is not equivalent.}
    \label{fig:logical-fail}
\end{figure}

\begin{figure}[t]
    \centering
    \includegraphics[width=0.4\textwidth]{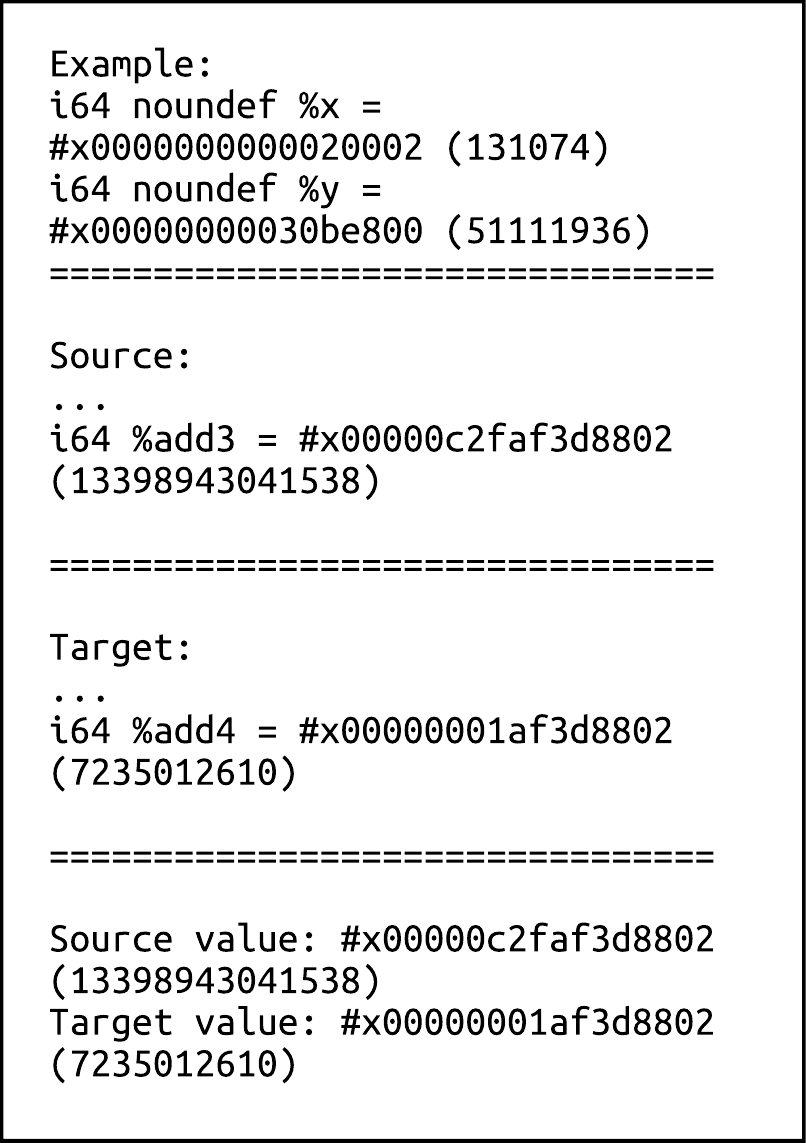}
    \caption{Formal equivalence validation failure: parameters \texttt{x} and \texttt{y} produce different outputs in the reference function and the variant.}
    \label{fig:eq-fail}
\end{figure}

The average rate of success decreases for test validation and equivalence verification to \numprint{82.6}\% (248/300) and \revised{\numprint{34.3}\%} (102/300), respectively.
Testing fails in 40 out of 288 remaining cases, either because of at least one failed test case or test suite timeout. 
An example from ffmpeg's \texttt{flac\_get\_max\_frame\_size} is shown in \ref{fig:logical-fail}, where constant addition to the \texttt{count} is not performed correctly.
Finally, equivalence validation fails in 151 out of 248 remaining cases.
Here, we uncover the cases where the LLM is able to synthesize variants that successfully run and behave similarly to the original, yet the complete semantics of the reference function are not correctly captured.
\autoref{fig:eq-fail} shows an equivalence validation failure example for one of the variants of function \texttt{fBlaMka}.
The figure shows a snippet of \texttt{alive-tv}'s that indicates that for inputs \texttt{x} and \texttt{y} of \numprint{131074} and \numprint{51111936} respectively, the reference function outputs \numprint{13398943041538}, while the variant outputs \numprint{7235012610}.

\begin{figure}[t]
    \centering
    \includegraphics[width=0.4\textwidth]{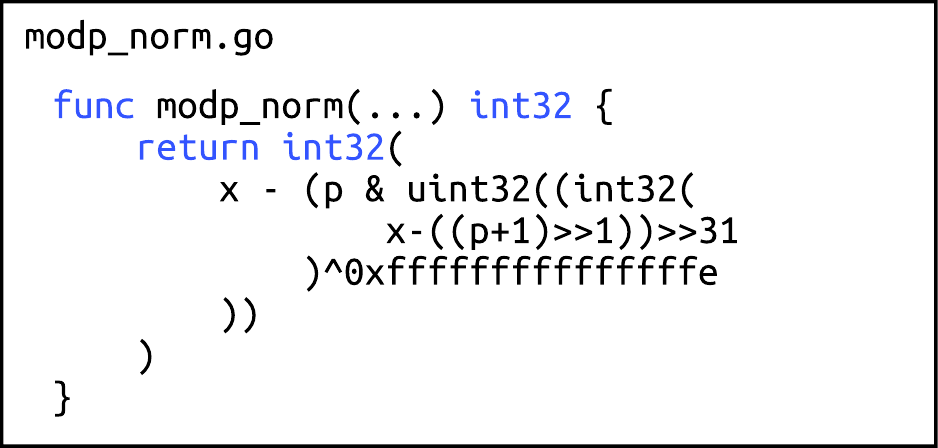}
    \caption{Build in isolation failure example. The code fails with an integer overflow error.}
    \label{fig:overflow-fail}
\end{figure}

For the cross-language configuration, on average, \numprint{92.6}\% (278/300) of the variants are compiled correctly in isolation, yet, when inserted in the project, only \numprint{76.3}\% (229/300) result in compilation success.
Compilation in isolation fails in 22 out of 300 cases, because of syntax errors and failure of compile-time checks, e.g. overflow checks.
\autoref{fig:overflow-fail} shows an example where the code fails to compile, as it fails to pass the go compiler's integer overflow check.

Compilation within the project fails in 49 out of 278 remaining cases, with the reasons behind these failures being (1) the use of IR components in the variant which are not declared in the project's IR code, and;
(2) IR type mismatches.
At this stage, an interesting case is openssl's \texttt{\_booth\_recode\_w7} function.
This function fails to compile within the project because the variants have a modified signature: The LLM added a suffix for each variant, disregarding the prompt's explicit instructions to keep the original signature.

When executing the variants, the average rate of success drops for test validation and equivalence verification to \numprint{69.0}\% (207/300) and \numprint{24.3}\% (73/300), respectively.
Testing fails in 22 out of 229 remaining cases, either because of at least one failed test case or test suite timeout.
The acquired errors are logical ones, such as the one displayed in \autoref{fig:logical-fail}, with no major LLM hallucinations detected in the variants generated in our experiments.
This is, no nonsensical, or completely off-the-mark solutions.
Finally, equivalence validation fails in 110 out of 207 remaining cases, where the LLM misses a part of the semantics of the reference function. 

It is worth noting that the acceptability of variants is close for both same-language and cross-language configurations, especially when contrasting the final number of variants that are proved to be equivalent.

Furthermore, for most functions (21/30), both configurations generate a similar amount of equivalent variants. Both same- and cross-language seem to excel or face difficulties for these functions: \texttt{libsodium}'s conversion functions are a clear example of this, consisting of a single line with several chained operations; they appear to \textit{confuse} the LLM, regardless of the diversification configuration.
The remaining functions (9/30) show disparate diversification results across configurations: while for some functions the same-language configuration performs better (5/9), for others the cross-language configuration produces more equivalent variants (4/9).

We now analyze the distribution of equivalent variants per function.
In the same-language configuration, for the bottom quartile, 0 equivalent variant were found;
for the median, 2 equivalent variants were found;
and the top quartile, 7 equivalent variants were found.
In the cross-language configuration, for the bottom quartile 0 equivalent variant were found;
for the median 1, equivalent variant was found; and for the top quartile, 5 equivalent variants.
We can observe that both distributions are skewed towards zero, this means that non-equivalent variants are more likely to be generated.

However, it is notable that for most (23/30) of the functions, at least 1 equivalent variant was generated.
We consider this to be a promising result that shows the feasibility of the \toolname concept. We believe this can be further improved by using more advanced models and sophisticated prompting techniques.

\begin{mdframed}[style=mpdframe,  frametitle=Answer to RQ1]
Large language models successfully generate variants given a reference function. Out of 600 function variants, 170 were verifiably equivalent, 97 of which were created from same-language source code and 73 from a source code in a different language.
This signals compelling applicability of creating automated cross-language N-Version functions with strong formal guarantees.
\end{mdframed}

\subsection{Static Uniqueness}
\label{sec:static-uniqueness}

\begin{table*}[]
    \centering
    \rowcolors{9}{gray!10}{white}
    \begin{tabular}{cl rrrrrr|rrrrrr|rr}
        \toprule
        \multicolumn{1}{c}{\multirow{4}{*}{\textbf{Project}}} & \multicolumn{1}{c}{\multirow{4}{*}{\textbf{Function Name}}} & \multicolumn{6}{c}{\textbf{\makecell{Same-language \\ diversification (C)}}} & \multicolumn{6}{c}{\textbf{\makecell{Cross-language \\ diversification (C-to-Go)}}} & \multicolumn{2}{c}{\textbf{Totals}} \\
        \cmidrule(lr){3-8} \cmidrule(lr){9-14} \cmidrule(lr){15-16}
        {} & {} & \multicolumn{1}{c}{\multirow{2}{*}{\footnotesize\textbf{\makecell{Total \\ eq.}}}} & \multicolumn{1}{c}{\multirow{2}{*}{\textbf{\footnotesize IR}}} & \multicolumn{4}{c}{\footnotesize \textbf{Binary}} & \multicolumn{1}{c}{\multirow{2}{*}{\footnotesize \textbf{\makecell{Total \\ eq.}}}} & \multicolumn{1}{c}{\multirow{2}{*}{\footnotesize \textbf{IR}}} & \multicolumn{4}{c}{\footnotesize \textbf{Binary}} & \multicolumn{1}{c}{\multirow{2}{*}{\footnotesize \textbf{IR}}} & \multicolumn{1}{c}{\multirow{2}{*}{\scriptsize \textbf{-O3}}} \\
        \cmidrule(lr){5-8} \cmidrule(lr){11-14}
         & & & & \scriptsize \textbf{-O0} & \scriptsize \textbf{-O1} & \scriptsize \textbf{-O2} & \scriptsize \textbf{-O3} & & & \scriptsize \textbf{-O0} & \scriptsize \textbf{-O1} & \scriptsize \textbf{-O2} & \scriptsize \textbf{-O3}\\
\midrule
\cellcolor{white} & \small alaw\_to\_s16 & \numprint{1} & \numprint{1} & \numprint{1} & \numprint{1} & \numprint{1} & \numprint{1} & \numprint{6} & \numprint{4} & \numprint{3} & \numprint{2} & \numprint{2} & \numprint{2} & \numprint{5} & \numprint{3} \\
\cellcolor{white} & \small iec958\_parity & \numprint{4} & \numprint{3} & \numprint{3} & \numprint{3} & \numprint{3} & \numprint{3} & \numprint{4} & \numprint{4} & \numprint{3} & \numprint{2} & \numprint{2} & \numprint{2} & \numprint{7} & \numprint{5} \\
\cellcolor{white} & \small add & \numprint{9} & \numprint{9} & \numprint{9} & \numprint{9} & \numprint{9} & \numprint{9} & \numprint{1} & \numprint{1} & \numprint{1} & \numprint{1} & \numprint{1} & \numprint{1} & \numprint{10} & \numprint{10} \\
\cellcolor{white} & \small ulaw\_to\_s16 & \numprint{9} & \numprint{9} & \numprint{8} & \numprint{8} & \numprint{8} & \numprint{8} & \numprint{10} & \numprint{2} & \numprint{1} & \numprint{1} & \numprint{1} & \numprint{1} & \numprint{11} & \numprint{9} \\
\cellcolor{white}
\multirow{-5}{*}{alsa-lib} & \small val\_seg & \numprint{2} & \numprint{2} & \numprint{2} & \numprint{2} & \numprint{2} & \numprint{2} & -- & -- & -- & -- & -- & -- & \numprint{2} & \numprint{2} \\
\midrule
\cellcolor{white} & \small flac\_get\_max\_frame\_size & \numprint{4} & \numprint{4} & \numprint{3} & \numprint{3} & \numprint{3} & \numprint{3} & -- & -- & -- & -- & -- & -- & \numprint{4} & \numprint{3} \\
\cellcolor{white} & \small mix & \numprint{6} & \numprint{6} & \numprint{5} & \numprint{5} & \numprint{5} & \numprint{5} & -- & -- & -- & -- & -- & -- & \numprint{6} & \numprint{5} \\
\cellcolor{white}
\multirow{-3}{*}{ffmpeg}  & \small weight & \numprint{10} & \numprint{9} & \numprint{8} & \numprint{8} & \numprint{8} & \numprint{8} & -- & -- & -- & -- & -- & -- & \numprint{9} & \numprint{8} \\

\midrule
\cellcolor{white} & \small barrett\_reduce & -- & -- & -- & -- & -- & --& \numprint{10} & \numprint{2} & \numprint{1} & \numprint{1} & \numprint{1} & \numprint{1} & \numprint{2} & \numprint{1} \\
\cellcolor{white} & \small ctz & \numprint{9} & \numprint{9} & \numprint{9} & \numprint{9} & \numprint{9} & \numprint{9} & \numprint{6} & \numprint{5} & \numprint{4} & \numprint{4} & \numprint{4} & \numprint{4} & \numprint{14} & \numprint{13} \\
\cellcolor{white} & \small \small int16\_t\_negative\_mask & \numprint{7} & \numprint{5} & \numprint{5} & \numprint{1} & \numprint{1} & \numprint{1} & \numprint{5} & \numprint{3} & \numprint{2} & \numprint{1} & \numprint{1} & \numprint{1} & \numprint{8} & \numprint{2} \\
\cellcolor{white} & \small int16\_t\_nonzero\_mask & \numprint{7} & \numprint{7} & \numprint{6} & \numprint{4} & \numprint{4} & \numprint{4} & \numprint{5} & \numprint{4} & \numprint{1} & \numprint{1} & \numprint{1} & \numprint{1} & \numprint{11} & \numprint{5} \\
\cellcolor{white}
\multirow{-5}{*}{libgcrypt} & \small montgomery\_reduce & \numprint{4} & \numprint{4} & \numprint{4} & \numprint{4} & \numprint{4} & \numprint{4} & \numprint{8} & \numprint{1} & \numprint{1} & \numprint{1} & \numprint{1} & \numprint{1} & \numprint{5} & \numprint{5} \\
\midrule
\cellcolor{white} & \small fpr\_half & \numprint{2} & \numprint{2} & \numprint{2} & \numprint{2} & \numprint{2} & \numprint{2} & \numprint{1} & \numprint{1} & \numprint{1} & \numprint{1} & \numprint{1} & \numprint{1} & \numprint{3} & \numprint{3} \\
\cellcolor{white} & \small fpr\_lt & -- & -- & -- & -- & -- & -- & \numprint{1} & \numprint{1} & \numprint{1} & \numprint{1} & \numprint{1} & \numprint{1} & \numprint{1} & \numprint{1} \\
\cellcolor{white} & \small modp\_montymul & \numprint{2} & \numprint{2} & \numprint{2} & \numprint{2} & \numprint{2} & \numprint{2} & -- & -- & -- & -- & -- & -- & \numprint{2} & \numprint{2} \\
\cellcolor{white} & \small modp\_norm & -- & -- & -- & -- & -- & -- & \numprint{2} & \numprint{1} & \numprint{1} & \numprint{1} & \numprint{1} & \numprint{1} & \numprint{1} & \numprint{1} \\
\cellcolor{white}
\multirow{-5}{*}{liboqs} & \small int16\_nonzero\_mask & \numprint{3} & \numprint{3} & \numprint{2} & \numprint{1} & \numprint{1} & \numprint{1} & \numprint{8} & \numprint{3} & \numprint{1} & \numprint{1} & \numprint{1} & \numprint{1} & \numprint{6} & \numprint{2} \\
\midrule
\cellcolor{white}
\multirow{-1}{*}{libsodium} & \small fBlaMka & \numprint{9} & \numprint{9} & \numprint{6} & \numprint{6} & \numprint{6} & \numprint{6} & -- & -- & -- & -- & -- & -- & \numprint{9} & \numprint{6} \\
\midrule
\cellcolor{white} & \small icbrt64 & \numprint{7} & \numprint{5} & \numprint{3} & \numprint{2} & \numprint{2} & \numprint{2} & -- & -- & -- & -- & -- & -- & \numprint{5} & \numprint{2} \\
\cellcolor{white} & \small BitDeinterleave & -- & -- & -- & -- & -- & -- & \numprint{3} & \numprint{2} & \numprint{2} & \numprint{2} & \numprint{2} & \numprint{2} & \numprint{2} & \numprint{2} \\
\cellcolor{white} & \small BitInterleave & \revised{1} & \revised{1} & \revised{1} & \revised{1} & \revised{1} & \revised{1} & \numprint{1} & \numprint{1} & \numprint{1} & \numprint{1} & \numprint{1} & \numprint{1} & \revised{\numprint{2}} & \revised{\numprint{2}} \\
\cellcolor{white}
\multirow{-4}{*}{openssl} & \small \_booth\_recode\_w5 & \numprint{2} & \numprint{2} & \numprint{2} & \numprint{2} & \numprint{2} & \numprint{2} & \numprint{2} & \numprint{2} & \numprint{1} & \numprint{1} & \numprint{1} & \numprint{1} & \numprint{4} & \numprint{3} \\
\midrule
Total & & \revised{98} & \revised{92} & \revised{81} & \revised{73} & \revised{73} & \revised{73} & 73 & 36 & 24 & 21 & 21 & 21 & \revised{128} & \revised{94} \\
\bottomrule
    \end{tabular}
    \caption{Variant uniqueness. Each column for both configurations shows how many unique variants are obtained at the IR level, and at the machine code level after various optimizations. Functions with no equivalent variants are omitted. The two rightmost columns display the total unique variants for both configurations, both at the IR level and at the machine code level, after building with the -O3 optimization flag.}
    \label{tab:static-uniqueness}
\end{table*}

\autoref{tab:static-uniqueness} shows the number of variants which compile to LLVM-unique IR code, and produce a unique machine code with different optimization flags.
The column "Total eq." shows the total number of equivalent variants found, as displayed in \autoref{tab:results-rq1}.
The column "IR" shows the number of these variants that are unique IR files. We obtain this number by comparing the "Total eq." variants' hashes at the IR level.
Subsequent columns, "-O0" to "-O3", show the number of unique machine code produced by the compiler with each optimization configuration. 

For the same-language configuration, 92 of the 98 equivalent variants are different from each other at the IR level. 
These 92 variants are generated from 19 out of the 30 reference functions. 
The compilation with "-O0" preserves 81 unique variants. Meanwhile, the number of unique variants is reduced to 73 after "-O3" optimization, representing 74.48\% of all equivalent variants.
The optimizations observed in the functions are performed at the "-O1" level, and are concerned with inlining, loop invariant code motion, and global value numbering.

For the cross-language configuration, a total of 36 unique variants were found at the IR level, among the 73 equivalent variants.

These 31 variants correspond to 16 out of 30 reference functions.
On compilation, the number of unique variants is reduced to 21 after "-O3" optimization, representing 28.76\% of all equivalent variants. 

A key result for RQ2 is the fact that the introduced diversity is generally well-preserved across optimization stages: from all 128 unique variants at the IR level, a total of 94 (73.43\%) are still unique after applying all optimization passes.

In total, at least 1 unique, equivalent variant was found for 24 out of 30 reference functions.
Furthermore, we are able to find at least one behaviorally unique variant for each function for which we successfully synthesize at least 1 equivalent variant.
In the case of same-language diversification, the number of unique variants per function is greater than 2 in most cases (15/19).
Cross-language diversification, on the other hand, does not perform as well, with only one unique variant for most functions (13/16).
We hypothesize that the lower number of unique variants in the cross-language configuration is caused by two reasons:
The LLM used in our experiment produces less diverse source code given our experimental parameters, or; 
the Go-to-IR compiler produces more consistent code across variants.

\subsection{Dynamic Uniqueness}
\label{sec:dynamic-uniqueness}

\begin{figure*}[h]
     \centering
     \begin{subfigure}[b]{0.32\textwidth}
         \centering
         \includegraphics[width=\textwidth]{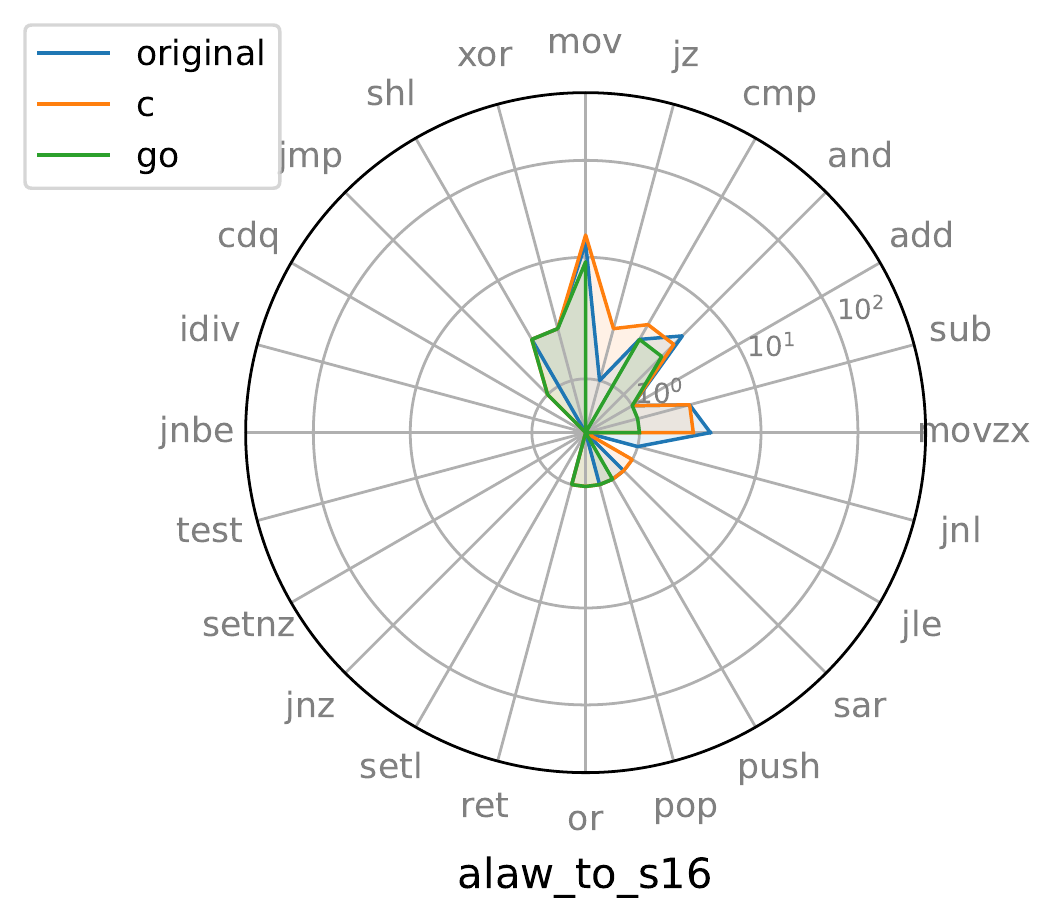}
     \end{subfigure}
     \begin{subfigure}[b]{0.32\textwidth}
         \centering
         \includegraphics[width=\textwidth]{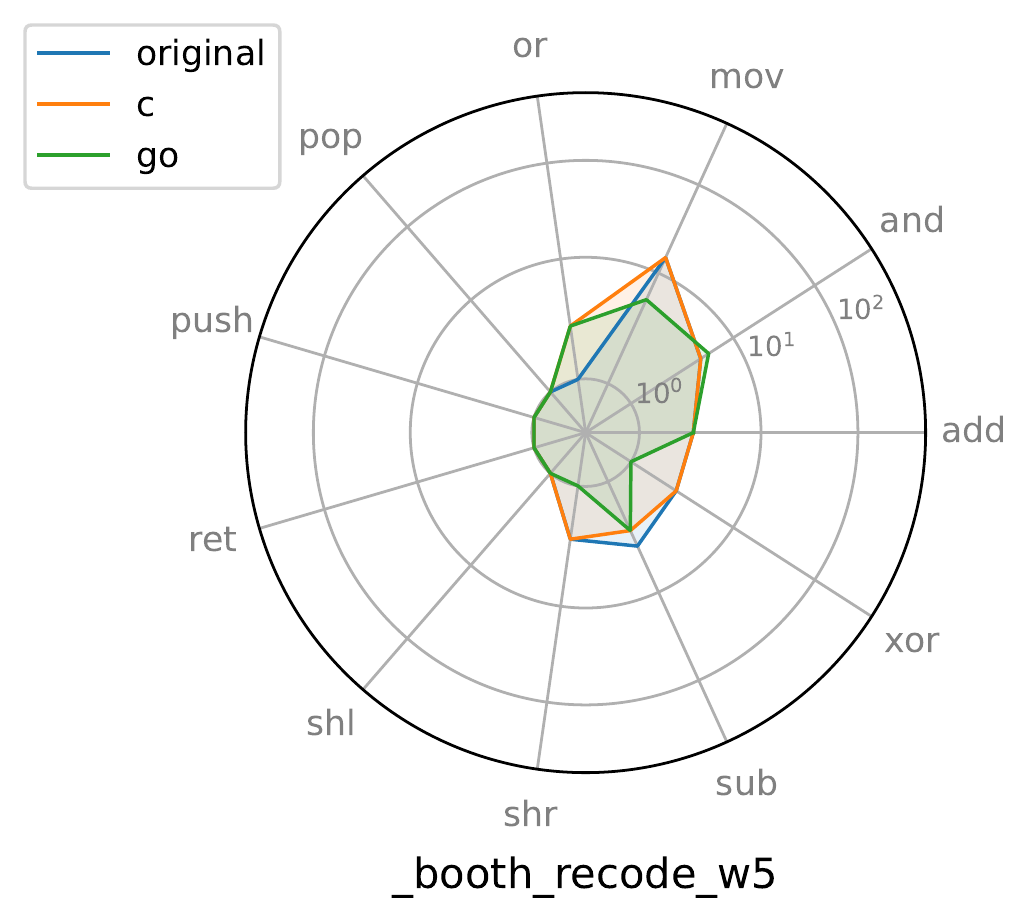}
     \end{subfigure}
     \begin{subfigure}[b]{0.32\textwidth}
         \centering
         \includegraphics[width=\textwidth]{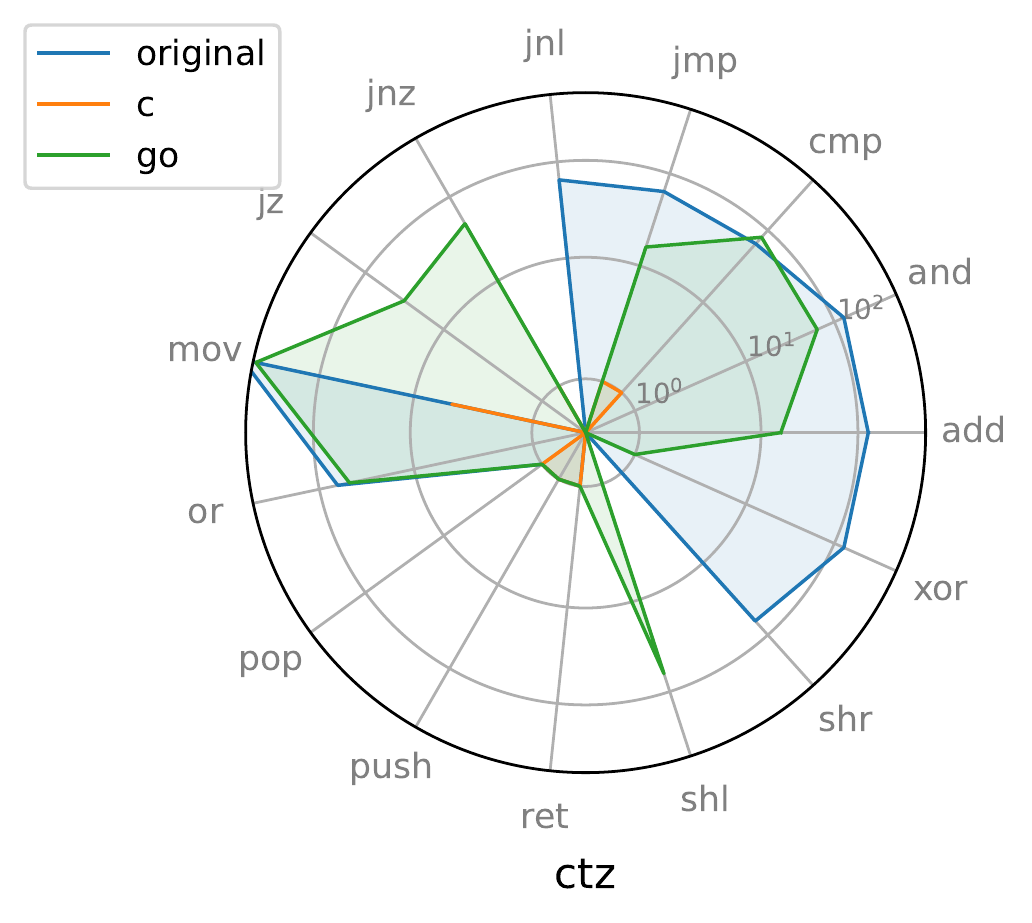}
     \end{subfigure}
     \begin{subfigure}[b]{0.32\textwidth}
         \centering
         \includegraphics[width=\textwidth]{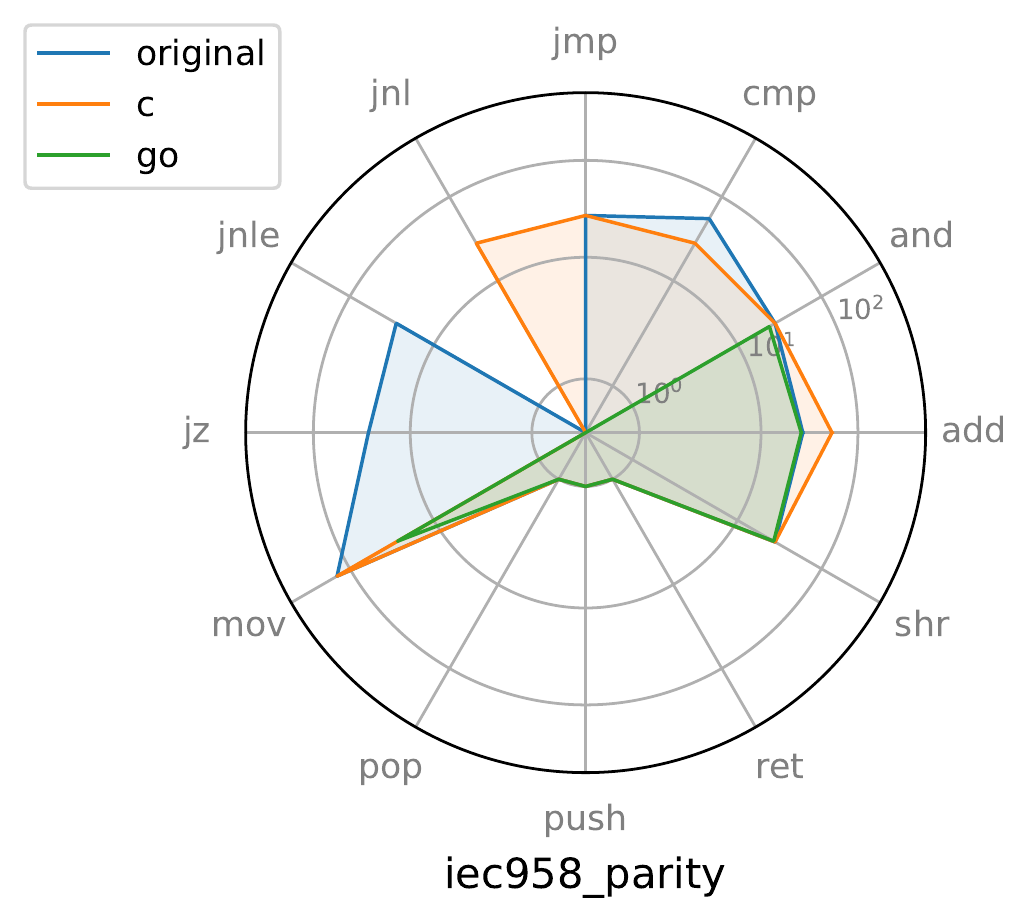}
     \end{subfigure}
     \begin{subfigure}[b]{0.32\textwidth}
         \centering
         \includegraphics[width=\textwidth]{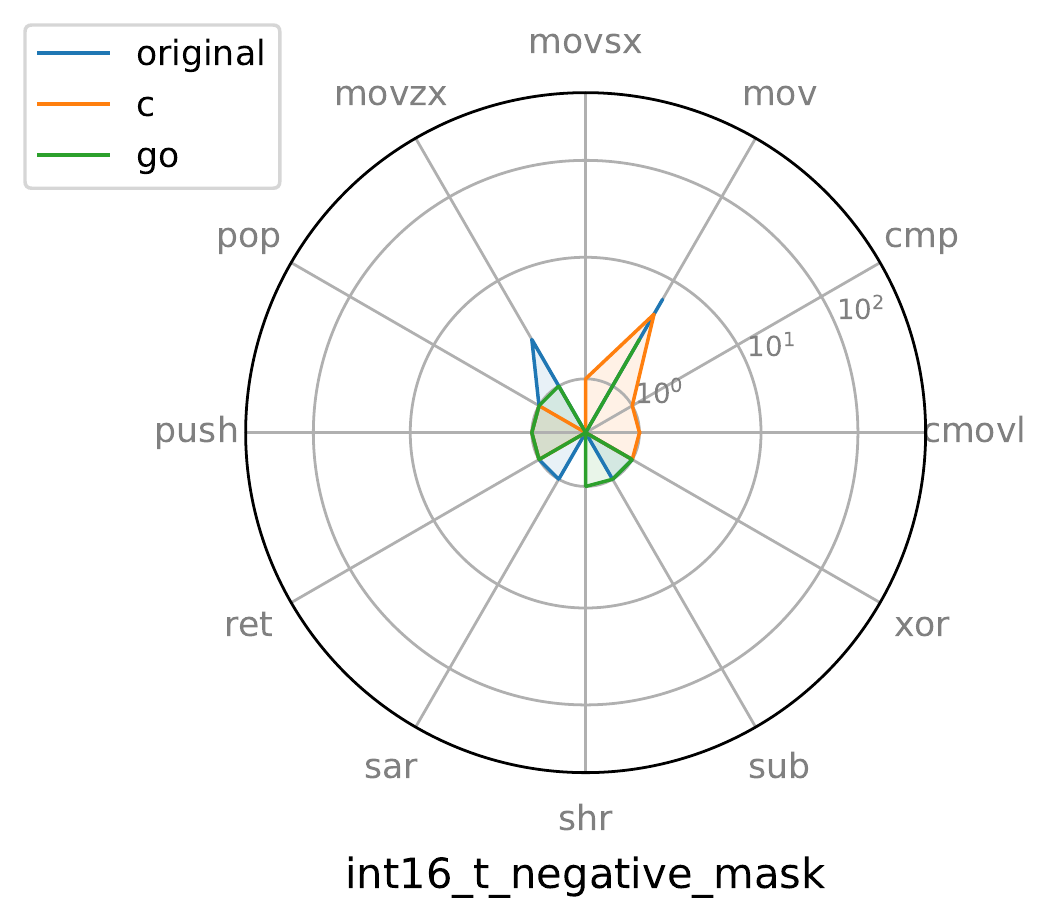}
     \end{subfigure}
     \begin{subfigure}[b]{0.32\textwidth}
         \centering
         \includegraphics[width=\textwidth]{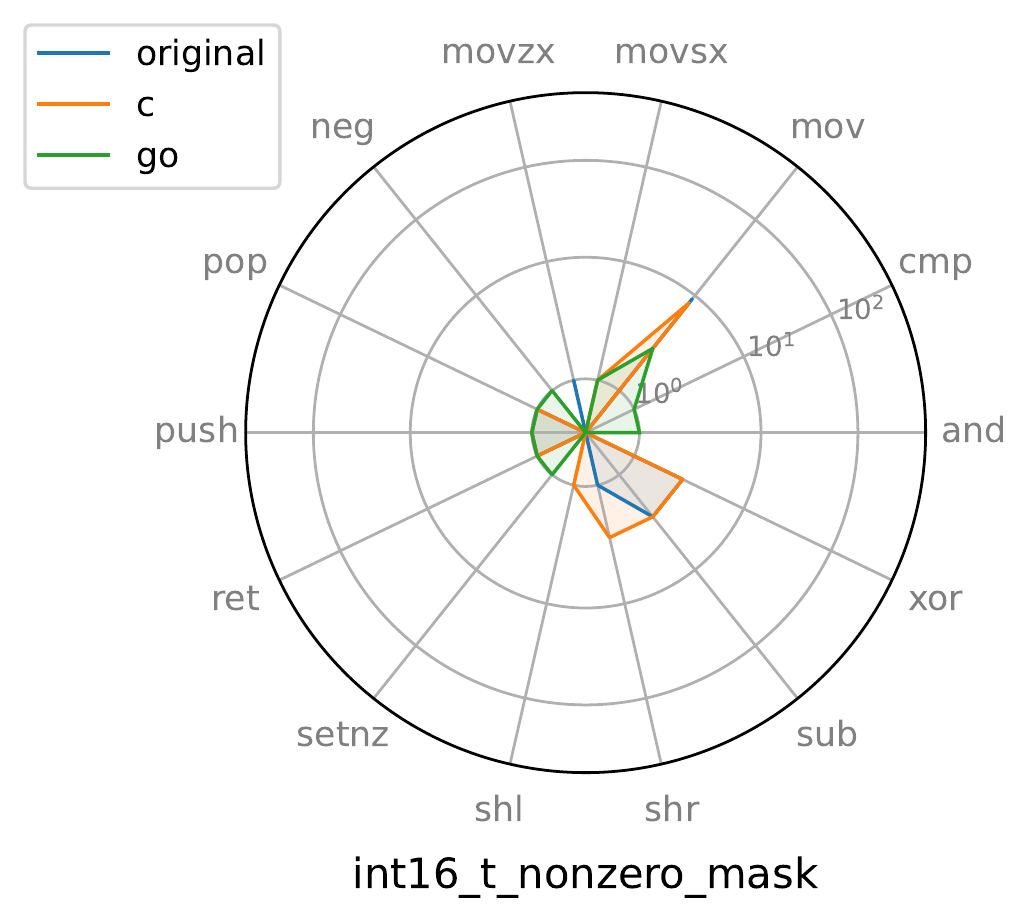}
     \end{subfigure}
\caption{Dynamic uniqueness of variants. Each radial plot shows the count of CPU instructions executed by the original function, superimposed with the instruction counts of the same-language (C) and cross-language (Go) variants.}
\label{fig:dynamic-uniqueness}
\end{figure*}

Now, we execute the diverse, equivalent variants generated by \toolname.
\autoref{fig:dynamic-uniqueness} shows plots with CPU instruction counts of 6 of the dataset's functions, one function per project.
Each sub-figure shows 3 overlapping areas, aggregating the instruction counts of the original function, an equivalent same-language variant, and an equivalent cross-language variant.
In detail, each figure is a radial plot, where the axes represent each of the CPU instructions executed by the function variants.
Each colored area aggregates the instruction count of a different function implementation.
This means that the less overlap in the figures, the more diverse the variants are in terms of the number and type of executed CPU instructions.

On these plots, we observe that the cross-language variants consistently show more diverse behavior in terms of CPU instructions. 
For instance, the sub-figure of function \texttt{\_booth\_recode\_w5} shows an almost complete overlap of the original implementation and the same-language variant's areas, while the cross-language variant area clearly displays a different set of instructions and count of the overlapping instructions.
Similarly, in the case of function \texttt{ctz}, the cross-language variant executes 3 instructions that are executed neither in the reference function nor in the same-language variant, while also avoiding 3 instructions from the original execution set.
This difference in instruction subsets follows naturally from using different compiler front-ends for producing the variants' IR code.

{\color{black}\begin{table*}[h]
    \centering
    \rowcolors{5}{gray!10}{white}
    \begin{tabular}{cl rrr|rrr|r}
        \toprule
        \multicolumn{1}{c}{\multirow{2}{*}{\color{black}\textbf{Project}}} & \multicolumn{1}{c}{\multirow{2}{*}{\color{black}\textbf{Function Name}}} & \multicolumn{3}{c}{\color{black}\textbf{C-to-C variants}} & \multicolumn{3}{c}{\color{black}\textbf{C-to-Go variants}} & \multicolumn{1}{c}{\multirow{2}{*}{\color{black}\textbf{\makecell{Tigress \\ variant}}}} \\
        \cmidrule(lr){3-5} \cmidrule(lr){6-8} 
        {} & {} & \color{black}\footnotesize\textbf{Min} & \color{black}\textbf{\footnotesize Median} & \color{black}\footnotesize \textbf{Max} & \color{black}\footnotesize \textbf{Min} & \color{black}\footnotesize \textbf{Median} & \color{black}\footnotesize \textbf{Max} & {} \\
\midrule
\cellcolor{white}  & \color{black}\small add & \color{black}0.538 & \color{black}0.700 & \color{black}0.800 & \color{black}\textbf{0.462} & \color{black}0.462 & \color{black}0.462 & \color{black}0.818 \\
\cellcolor{white}  & \color{black}\small iec958\_parity & \color{black}0.750 & \color{black}0.750 & \color{black}0.833 & \color{black}\textbf{0.500} & \color{black}0.636 & \color{black}0.636 & \color{black}0.733 \\
\cellcolor{white}  \multirow{-3}{*}{\color{black}alsa-lib} & \color{black}\small val\_seg & \color{black}0.727 & \color{black}1.000 & \color{black}1.000 & \color{black}-- & \color{black}-- & \color{black}-- & \color{black}\textbf{0.600} \\
\midrule
\cellcolor{white}  & \color{black}\small flac\_get\_max\_frame\_size & \color{black}1.000 & \color{black}1.000 & \color{black}1.000 & \color{black}-- & \color{black}-- & \color{black}-- & \color{black}\textbf{0.688} \\
\cellcolor{white} \multirow{-2}{*}{\color{black}ffmpeg}  & \small \color{black}mix & \color{black}1.000 & \color{black}1.000 & \color{black}1.000 & \color{black}-- & \color{black}-- & \color{black}-- & \color{black}\textbf{0.714} \\
\midrule
\cellcolor{white}  & \color{black}\small barrett\_reduce & \color{black}-- & \color{black}-- & \color{black}-- & \color{black}0.900 & \color{black}0.900 & \color{black}0.900 & \color{black}\textbf{0.769} \\
\cellcolor{white}  & \color{black}\small ctz & \color{black}0.462 & \color{black}0.462 & \color{black}0.643 & \color{black}\textbf{0.444} & \color{black}0.538 & \color{black}0.667 & \color{black}0.750 \\
\cellcolor{white}  & \color{black}\small int16\_t\_negative\_mask & \color{black}\textbf{0.417} & \color{black}0.417 & \color{black}0.700 & \color{black}0.556 & \color{black}0.778 & \color{black}0.778 & \color{black}0.727 \\
\cellcolor{white}  & \color{black}\small int16\_t\_nonzero\_mask & \color{black}0.357 & \color{black}0.455 & \color{black}0.700 & \color{black}\textbf{0.308} & \color{black}0.308 & \color{black}0.308 & \color{black}0.727 \\
\cellcolor{white}  \multirow{-5}{*}{\color{black}libgcrypt} & \color{black}\small montgomery\_reduce & \color{black}0.800 & \color{black}0.800 & \color{black}0.900 & \color{black}\textbf{0.636} & \color{black}0.636 & \color{black}0.636 & \color{black}0.750 \\
\midrule
\cellcolor{white}  & \color{black}\small fpr\_half & \color{black}\textbf{0.727} & \color{black}0.727 & \color{black}0.727 & \color{black}0.875 & \color{black}0.875 & \color{black}0.875 &\color{black} \textbf{0.727} \\
\cellcolor{white}  & \color{black}\small fpr\_lt & \color{black}-- & \color{black}-- & \color{black}-- & \color{black}\textbf{0.412} & \color{black}0.412 & \color{black}0.412 & \color{black}0.750 \\
\cellcolor{white}  & \color{black}\small int16\_nonzero\_mask & \color{black}0.357 & \color{black}0.417 & \color{black}0.417 & \color{black}\textbf{0.286} & \color{black}0.286 & \color{black}0.286 & \color{black}0.750 \\
\cellcolor{white}  & \color{black}\small modp\_montymul & \color{black}\textbf{0.750} & \color{black}0.750 & \color{black}1.000 & \color{black}-- & \color{black}-- & \color{black}-- & \color{black}0.769 \\
\cellcolor{white}  \multirow{-5}{*}{\color{black}liboqs} & \color{black}\small modp\_norm & \color{black}-- & \color{black}-- & \color{black}-- & \color{black}1.000 & \color{black}1.000 & \color{black}1.000 & \color{black}\textbf{0.727} \\
\midrule
\cellcolor{white}  \multirow{-1}{*}{\color{black}libsodium} & \color{black}\small fBlaMka & \color{black}0.875 & \color{black}1.000 & \color{black}1.000 & \color{black}-- & \color{black}-- & \color{black}-- & \color{black}\textbf{0.667} \\
\midrule
\cellcolor{white}  & \color{black}\small BitDeinterleave & \color{black}-- & \color{black}-- & \color{black}-- & \color{black}\textbf{0.364} & \color{black}0.500 & \color{black}0.500 & \color{black}0.571 \\
\cellcolor{white}  & \color{black}\small BitInterleave & \color{black}\textbf{0.333} & \color{black}0.333 & \color{black}0.333 & \color{black}0.500 & \color{black}0.500 & \color{black}0.500 & \color{black}0.571 \\
\cellcolor{white} \multirow{-3}{*}{\color{black}openssl} & \color{black}\small \_booth\_recode\_w5 & \color{black}1.000 & \color{black}1.000 & \color{black}1.000 & \color{black}1.000 & \color{black}1.000 & \color{black}1.000 & \color{black}\textbf{0.786} \\
\bottomrule
\end{tabular}
\caption{
\color{black}Jaccard coefficient comparison of executed x86 instructions to quantify runtime diversity 
introduced by {\toolname} and Tigress' flatten transform. 
For each function, we list the minimum, median, and maximum Jaccard coefficients for the C-to-C and C-to-Go variants produced by {\toolname}, as well as the coefficient for the variant generated by Tigress. 
Lower coefficients indicate less overlap with the reference implementation and thus greater runtime diversity. 
Bold values indicate the greatest diversity (lowest coefficient) observed for each function. 
Missing entries (--) correspond to configurations where no equivalent variant could be produced.
}
\label{tab:dynamic-uniqueness-comparison}
\end{table*}}

\revised{
Finally, Table~\ref{tab:dynamic-uniqueness-comparison} reports the overlap between the variants and their reference implementations, measured with the Jaccard coefficient over executed x86 instructions.
Across 12 of the 19 functions, at least one variant produced by \toolname attains a lower Jaccard coefficient than the variant produced by Tigress' flatten transform.
Overall, {\toolname}'s variants exhibit lower coefficients on average than Tigress', indicating less overlap and therefore greater runtime diversity.
Moreover, the minimum observed Jaccard coefficient for \toolname is 0.286, compared to 0.636 for Tigress' flatten transform.
Lower coefficients imply smaller overlaps in executed instructions, i.e., greater diversity at runtime.

Focusing on {\toolname}'s variants, among the 11 functions where both same-language (C-to-C) and cross-language (C-to-Go) equivalents were found, 8 have their lowest coefficient coming from a cross-language variant.
This suggests that cross-language variants are generally more effective at maximizing runtime diversity.
}

\begin{mdframed}[style=mpdframe, frametitle=Answer to RQ2]
Our experiments show that \toolname is able to create function variants that are diverse to a large extent, as measured through their persistence after compilation and optimization. 
In total, for 166 unique variants across the 30 reference functions' source code, 126 (75.9\%) remained unique after all compilation and optimization passes. 
This demonstrates that \toolname's approach to diversity is robust to different optimization requirements.

\revised{We also compared \toolname to a state-of-the-art rule-based diversification tool, Tigress. 
While Tigress is able to generate variants with runtime diversity, our results show that \toolname generally achieves greater runtime diversity in the executed x86 instruction sequences. 
For 12/19 functions, the variants produced by \toolname exhibit lower Jaccard coefficients relative to the reference implementation than those produced by Tigress, indicating less overlap and thus greater runtime diversity. 
This suggests that \toolname's translation-based approach can introduce a higher degree of behavioral diversity than traditional rule-based diversification transformations. 
}

In general for the generated variants, we observe diverse internal behavior at runtime, evidenced by different sequences of CPU instructions.
This finding is relevant, as it means that the introduced diversity at the machine code level goes beyond minor changes in the binary.
It is aligned with the core assumption of N-Version programming, that the fault-tolerance increases with runtime diversity.

\end{mdframed}

\subsection{Mitigating Miscompilations}
\label{sec:mitigation}

\autoref{fig:miscompilation-results} shows the results of mitigating the Clang miscompilation bugs introduced in \autoref{fig:miscompilations} with \toolname.
First, we observe that \toolname successfully generates multiple equivalent variants for each function that triggers each miscompilation bug.
Second, \toolname is able to assemble N-Version binaries for these functions, which caused the execution to crash, instead of producing a wrong output.
We consider \toolname effective in mitigating the three proposed bugs.

\begin{figure*}[t]

\begin{subfigure}[b]{0.28\textwidth}
    \includegraphics[width=\textwidth]{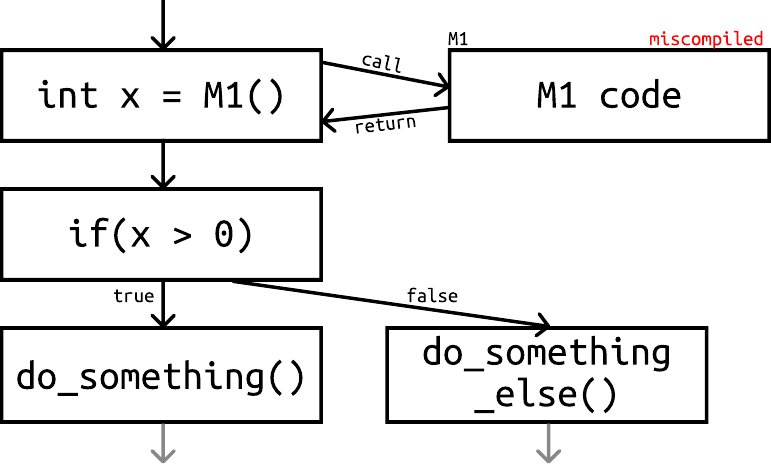}
\end{subfigure}
\hfill
\begin{subfigure}[b]{0.66\textwidth}
    \includegraphics[width=\textwidth]{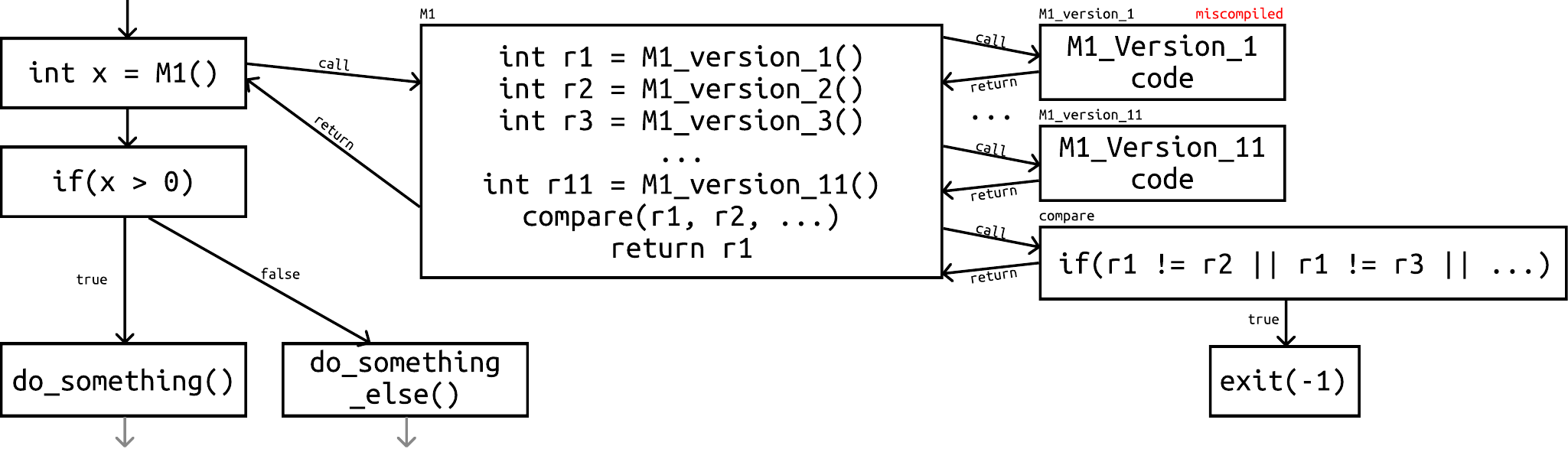}
\end{subfigure}

    \caption{Control flow graphs for M1. Left: single-version implementation. Right: N-Version implementation.}
    \label{fig:control-flow}
\end{figure*}

For bug M1, \toolname produces 10 equivalent variants in the same-language configuration.
It then assembles an 11-Version implementation of the function which triggers the miscompilation fault. 
The key result here is in the ability of the 11-Version function to let developers know that something wrong has happened.
The original function compiles and, when it executes, produces a value.
However, the value is wrong because of the miscompilation bug, but there is no warning or check to let the developers know about this wrong value.
On the other hand, the 11-Version function compiles, and when it executes, it crashes because the results returned by each version are not consistent.
In this case, the program does not silently introduce a wrong value and the developer can act upon this.
Consider the control flow graphs in \autoref{fig:control-flow}.
The graph on the left represents the execution of a single-version implementation of the M1 function.
The miscompiled code could result in the program silently choosing the wrong execution path, with potentially dangerous results.
The graph on the right represents the execution of an N-Version implementation of the M1 function.
The miscompiled code executes and returns silently; however, if at least any of the other 10 versions returns a distinct value, the program will exit, thus avoiding wrong execution and informing the program operator that something wrong has occurred.
This behavior is similar for the N-Version implementations that \toolname produced for M2 and M3, with N values of 11 and 3, respectively.

\begin{table}[h!]
    \centering
    \rowcolors{2}{gray!10}{white}
    \begin{tabular}{lllr}
    \toprule
    Bug & Result & Configuration & \#Eq. variants \# \\
    \midrule
    M1 & Mitigated & Same-language & \numprint{10} \\
    M2 & Mitigated & Same-language & \numprint{10} \\
    M3 & Mitigated & Same-language & \numprint{2} \\ 
    \bottomrule
         
    \end{tabular}
    \caption{\toolname' miscompilation mitigation results.}
    \label{fig:miscompilation-results}
\end{table}

Upon manual inspection of the execution of the resulting binaries, we find that none of the equivalent variants are affected by the miscompilation bugs, only the original function is.
As a result, the program termination is triggered by the inconsistent return values between the original function and the return values of one of the corresponding variants.

\begin{figure*}[t]
    \centering
    \includegraphics[width=0.9\textwidth]{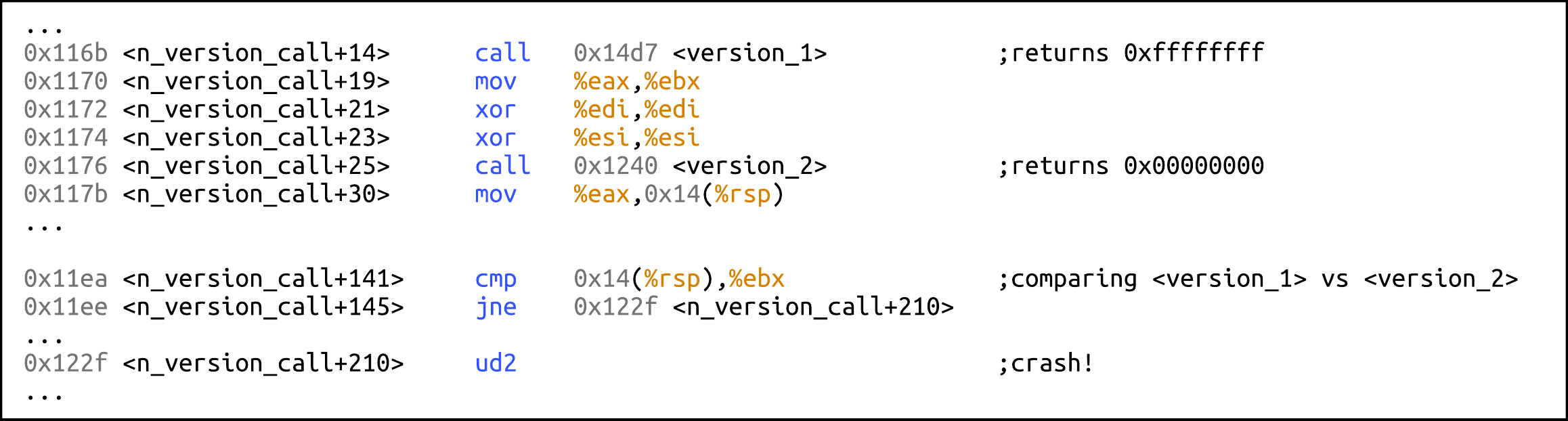}
    \caption{Disassembled machine code for the hardened M2 function.}
    \label{fig:bug-2-exec}
\end{figure*}

\autoref{fig:bug-2-exec} shows an excerpt of the machine code produced by \toolname, when attempting mitigation of M2.
The figure shows the calls to the first two versions of the resulting 11-Version implementation: \texttt{<version\_1>} is the original function, which is miscompiled; and \texttt{<version\_2>} is one of the variants proven to be equivalent at the IR level.
In this example, \texttt{<version\_1>} returns \texttt{-1}, while \texttt{<version\_2>} returns \texttt{0}.
Later in the execution, the return values are compared, and since these are different, the execution is redirected to the \texttt{ud2} instruction, triggering a crash.

\begin{mdframed}[style=mpdframe,frametitle=Answer to RQ3]
\toolname is able to generate N-Version programs that mitigate the considered real-world miscompilation bugs.
Our results show that equivalent function variants produced by LLMs can be employed to harden critical sections of a given program by assembling N-Version functions.

\end{mdframed}

\subsection{Performance Overhead}
\autoref{tab:overhead}~shows a summary of the overhead introduced by the N-Version functions when compared to the reference functions.
The column "Baseline Runtime" shows the measurement taken for each function as described in \autoref{sec:rq4-methodology}.
For both same-language and cross-language configurations, "N" represents the number of versions used in the maximally diverse N-Version function; "N-Version runtime" shows the execution time for the N-Version function; and finally "Normalized Overhead" shows the average overhead for a N-Version function with one single-variant (i.e. a 2-version function).

The average normalized overhead is expected to be 2x for a 2-version function, which is what we observe on the last average row.
The observed maximum and minimum overheads are 3.75x and 1.04x respectively.
The minimum value can be considered an outlier, and it is introduced by a loop-unroll compiler optimization in the Go variant of a C function.
We note that it is also possible to design parallel execution harnesses in time-critical contexts.

\begin{mdframed}[style=mpdframe, frametitle=Answer to RQ4]
Our experiments show that the N-Version functions generated by \toolname introduce the expected overhead.
After executing the N-Version functions, we observe that the runtime is increased in average by 2x for a 2-version function, and grows linearly by adding more variants to the assembly. 
This experiment provides evidence that \toolname's harness pass' overhead is negligible.
We note that variants may introduce algorithmic differences in the diversified function (e.g. iteration vs recursion) changing the diversified function complexity itself, yet this does not change the overhead due to the N-Version assembly itself.
\end{mdframed}

\begin{table*}[]
    \centering
    \rowcolors{5}{gray!10}{white}
    \begin{tabular}{lrrrrrrr}
        \toprule
        \multicolumn{1}{c}{\multirow{3}{*}{\textbf{Function Name}}} & \multicolumn{1}{c}{\multirow{3}{*}{\textbf{\makecell{Baseline \\ runtime ($\mu$s)}}}} & \multicolumn{3}{c}{\textbf{\makecell{Same-language \\ diversification (C)}}} & \multicolumn{3}{c}{\textbf{\makecell{Cross-language \\ diversification (C+Go)}}}  \\
        \cmidrule(lr){3-5} \cmidrule(lr){6-8}
        {} & {} & \small \textbf{N} & \small \textbf{\makecell{N-Version \\ runtime ($\mu$s)}} & \small \textbf{\makecell{Normalized \\ Overhead}} & \small \textbf{N} & \small \textbf{\makecell{N-Version \\ runtime ($\mu$s)}} & \small \textbf{\makecell{Normalized \\ Overhead}} \\
        \midrule
alaw\_to\_s16 & \numprint{4606} & 2 & \numprint{17256} & 3.75x & 7 & \numprint{38363} & 2.22x\\
iec958\_parity & \numprint{83623} & 5 & \numprint{269676} & 1.56x & 5 & \numprint{95393} & 1.04x\\
add & \numprint{2486} & 10 & \numprint{36362} & 2.51x & 2 & \numprint{5879} & 2.36x\\
ulaw\_to\_s16 & \numprint{4369} & 10 & \numprint{54707} & 2.28x & 11 & \numprint{52961} & 2.11x\\
val\_seg & \numprint{4163} & 3 & \numprint{12544} & 2.01x & -- & -- & -- \\
flac\_get\_max\_frame\_size & \numprint{6222} & 5 & \numprint{30622} & 1.98x & -- & -- & -- \\
mix & \numprint{9067} & 7 & \numprint{60226} & 1.94x & -- & -- & -- \\
weight & \numprint{6256} & 11 & \numprint{59428} & 1.85x & -- & -- & -- \\
icbrt64 & \numprint{64412} & 8 & \numprint{700098} & 2.41x & -- & -- & -- \\
BitDeinterleave & \numprint{1745} & -- & -- & -- &4 & \numprint{7309} & 2.06x\\
BitInterleave & \numprint{1848} & -- & -- & -- &2 & \numprint{4293} & 2.32x\\
\_booth\_recode\_w5 & \numprint{4599} & 3 & \numprint{13757} & 2.0x & 3 & \numprint{11070} & 1.7x\\
fpr\_half & \numprint{2895} & 3 & \numprint{12139} & 2.6x & 2 & \numprint{5833} & 2.01x\\
fpr\_lt & \numprint{5615} & -- & -- & -- &2 & \numprint{13796} & 2.46x\\
modp\_montymul & \numprint{5543} & 3 & \numprint{20754} & 2.37x & -- & -- & -- \\
modp\_norm & \numprint{2793} & -- & -- & -- &3 & \numprint{8210} & 1.97x\\
int16\_nonzero\_mask & \numprint{3872} & 4 & \numprint{15601} & 2.01x & 9 & \numprint{25391} & 1.69x\\
barrett\_reduce & \numprint{3095} & -- & -- & -- &11 & \numprint{67948} & 3.1x\\
ctz & \numprint{212299} & 10 & \numprint{866981} & 1.34x & 7 & \numprint{655523} & 1.35x\\
int16\_t\_negative\_mask & \numprint{2754} & 8 & \numprint{25538} & 2.18x & 6 & \numprint{16344} & 1.99x\\
int16\_t\_nonzero\_mask & \numprint{4060} & 8 & \numprint{27467} & 1.82x & 6 & \numprint{18922} & 1.73x\\
montgomery\_reduce & \numprint{2472} & 5 & \numprint{18241} & 2.59x & 9 & \numprint{17304} & 1.75x\\
fBlaMka & \numprint{3356} & 10 & \numprint{38936} & 2.18x & -- & -- & -- \\

\midrule
Avg. Overhead & & & & 2.18x & & & 1.99x \\
\bottomrule
    \end{tabular}
    \caption{Runtime measurement of N-Version functions.}
    \label{tab:overhead}
\end{table*}
\label{sec:overhead}

\section{Discussion}

In this section, we discuss \toolname in the larger context of software development, as well as the threats to the validity of our experiments.

\subsection{Integration into Existing Software Development Practices}
The integration of \toolname into an existing software development project requires:
(1) identifying critical sections of code with the highest reliability standards,
and;
(2) isolating them into pure functions.
Developers would need to integrate \toolname in the build pipeline for those functions.
The pipeline would automatically generate variants, validate their correctness, and insert the N-Version functions into the final build.
At deployment time, the N-Version functions would terminate execution in case of inconsistencies detected by diversification.
This is valid in IoT \cite{almohri2022dynamic}.
We are also aware that companies providing security critical software for the military have such build pipelines with other kinds of diversification.

\subsection{Threats to Validity}

\textit{Threats to internal validity:}
Per our design, we rely on existing intermediate representations (LLVM IR) and equivalence checkers (alive-tv).
Therefore, \toolname will inherit their trustworthiness and limitations.
For instance, in our experiments, some of the variants soundly discarded as non-equivalent were not proven as such, but \texttt{alive-tv} failed to process them because of unsupported LLVM instructions.

Also, the inherent non-determinism of LLMs can affect our results.
This means that the proportion of equivalent variants can be different in an experimental reproduction.
However, non-determinism is key in the design of \toolname, as it is needed in the diversification pass to generate distinct variants.

\textit{Threats to external validity:}
We identify two dimensions where the obtained results can be generalized for \toolname' design.
First, our experiments focus on two language pairs: C-to-C and C-to-Go.
We argue that similar results can be achieved for other language pairs,
as long as these are IR-compatible, and a corresponding IR equivalence checking tool exists.
Second, our experiments only consider a single LLM as a source of diversity.
We argue that similar results can be achieved using different LLMs, as these have been shown to have similar capabilities~\cite{los-alamos}.
We acknowledge that the generalizability of the results, and the claims of suitability of the design can be strengthened by performing experiments with different language pairs, IRs, generative models, and prompt templates.

\section{Related Work}

\subsection{Automated Variant Synthesis}
Singh~\etal~\cite{avatars} propose the automatic creation of language-diverse, formally verified program variants from a high-level specification.
They focus on fault-tolerance of distributed systems by executing each variant following the replicated state machine paradigm.
However, no prototype implementation, experimental methodology, or results are provided.
Pelofske~\etal~\cite{los-alamos} present a thorough analysis of function variant generation with LLMs.
This study is focused on four SHA-1 functions, and provides a deep insight into the diversity and correctness of LLM-generated variant functions.
Xu~\etal~\cite{n-version-obf} describe a variant generation mechanism based on LLVM-IR obfuscation. 
Their goal is to create explicitly non-equivalent variants, which are resistant to replication of known tampering mechanisms.
While these works aim at automating variant creation, and to different extents, correctness verification, the key novelty of our work is the focus on hardening of binaries via N-Version function execution.

\subsection{LLMs code correctness}
Pan~\etal~\cite{survey-bugs} identify the correctness of LLM-generated code as an important challenge to address.
In this vein, Kessel and Atkinson~\cite{Kessel_2024} present LASSO, a platform for differential execution of large sets of code \textit{versions} and tests.
Compared to this work \toolname goes one step further and automates the creation of N-Version programs, in addition to generation and testing of code variants. 
Du~\etal~\cite{mercury} present a benchmark for solutions to programming tasks via LLMs.
This benchmark is designed to measure how efficient are the generated solutions in terms of algorithmic complexity.
The experiments show that LLMs are able to produce code that solves the tasks while having diverse efficiency scores.
Zhang~\etal~\cite{NEURIPS2023_43e9d647} introduce a benchmark to quantify the correctness of code produced by an LLM.
The experiments evidence that LLMs have the capability to correctly solve programming tasks to a large extent.
However, while the data provided by these works gives a detailed overview of the correctness and efficiency of code produced by LLMs, the evaluations are test-based, and not formally verified against a specification.

\subsection{Code Generation and Code Translation}

Code generation and translation with LLMs are very active research areas~\cite{survey1, survey2}.
Yang~\etal~\cite{vert} present a C-to-Rust verified translation tool.
It works by using static methods to construct a known equivalent translation, and LLMs to create more idiomatic translations.
The latter are then checked for equivalence against the former using an off-the-shelf Rust tool.
Bathia~\etal~\cite{llmlift} describe a tool to produce verified code transpilations.
This is achieved by leveraging intermediate representations and formal equivalence checkers.
Yang~\etal~\cite{unitrans} present UniTrans, a framework to generate and test LLM code translation.
UniTrans uses the original code to further generate test cases for the resulting translation, and leverage them to iteratively repair the translation in case of failure. 
While the mentioned works address the issue of validating correctness of LLM-generated code equivalence, their scope is not concerned with software reliability.

\subsection{Code Transplantation and Recombination:}
As part of \toolname' Harnessing pass, IR code from newly generated programs is inserted into the project that is being hardened.
This process is similar to the code transplantation mechanisms described in the literature.
This process has been studied in the context of automated program repair~\cite{semgraft}, feature transfer between systems~\cite{carbon-copy, plastic-surgery, babel-pidgin}, and code clone analysis~\cite{code-clones}.
The Harnessing pass is also similar to recombination approaches, where different versions of systems are created by manipulating sections of readily available codebases, at lower-than-source levels~\cite{frankenssl, object-level-recombination}.
While closely related, these works are not concerned with the challenges of the two first passes of \toolname, namely, automatic diversification and correctness checking.

\subsection{Addressing Miscompilations}
The works of Eide~\etal~\cite{regehr-volatile} and Le~\etal~\cite{miscomp-guided} present approaches to discover miscompilation bugs via stochastic exploration of the program space.
With the same goal, Tu~\etal~\cite{miscomp-llm} describe a mechanism with guided LLM prompt production approach, which allows producing bug-triggering program mutations.
While these approaches seek to actively discover miscompilation bugs, \toolname takes a program hardening, as it does not target compilers directly, but protects critical application code against potential miscompilation.

\section{Future Work}

\revised{Looking ahead, we see concrete directions opened by this research.}

\subsection{Independence of LLM-Generated Variants}

\revised{A fundamental challenge in N-Version programming is ensuring that different implementations fail independently under the same inputs. While our current approach generates variants that are statically and dynamically different, we need to explore how to measure independence of implementations. This includes defining quantitative metrics for independence, establishing experimental protocols to estimate fault correlation under diverse inputs.
Then, one could use these metrics to synthesize variants that actively maximize independence. Future work should also investigate whether LLM-generated variants exhibit different failure patterns, not only between themselves, but compared to human-written implementations.}

\subsection{LLM Prompting and Search Approaches}

\revised{The current diversification approach uses a single-shot prompting strategy. Future work should explore more sophisticated approaches to variant generation. This includes casting variant synthesis as constrained search that maximizes behavioral diversity under an equivalence oracle, and exploiting verifier counterexamples to repair failing variants. We should also investigate measurement-driven synthesis, where prompts are conditioned on diversity metrics (e.g., edit distance, instruction-set Jaccard) to steer generation away from previously explored regions. Additionally, diversity-aware fine-tuning could adapt LLMs to the diversification task, with training objectives that encourage behavioral distance while preserving equivalence.}

\subsection{Language Support Extension}

\revised{While our current implementation focuses on C and Go, extending \toolname to support additional programming languages presents both opportunities and challenges. The primary technical barrier is the requirement for IR compatibility—languages must compile to a common intermediate representation (e.g., LLVM IR) to enable formal equivalence checking. This constraint limits support to languages with mature LLVM front-ends, such as Rust or Swift. The implementation cost for adding a new language pair involves: (1) ensuring robust compilation to LLVM IR, (2) adapting the validation pipeline to handle language-specific IR patterns, and (3) potentially developing language-specific equivalence checking tools if existing ones are insufficient. However, the modular design of \toolname's three-pass architecture means that language support can be added incrementally without major architectural changes. The diversification pass is already language-agnostic, requiring only appropriate prompting strategies. The validation pass depends on IR-level tools that can be extended or replaced as needed. The harnessing pass operates at the IR level and is largely language-independent. Thus, while each new language pair requires careful engineering effort, the overall approach scales well to additional languages.}

\section{Conclusion}

\revised{We introduced automated N-Version programming powered by LLMs and presented \toolname, a practical pipeline that (1) diversifies functions at the source level, (2) validates candidates via compilation, testing, and formal IR-level equivalence checking, and (3) harnesses verified variants into executable N-Version software.

Across 30 real-world C functions, we found that LLMs can synthesize function variants that are provably equivalent, even across languages.
We showed that a substantial portion of the induced diversity survives compiler optimizations and can be measured at runtime.
We also demonstrated that assembling these variants as N-Version functions mitigates real miscompilation bugs, by converting silent wrong results into explicit program halts.
The added runtime cost matches expectations ($\sim$2x for a 2-version function) and scales linearly with N.
Overall, our results indicate that combining practical formal methods with modern LLMs can substantially reduce the engineering cost of N-Version programming while giving significant correctness guarantees.
}

\balance
\bibliographystyle{unsrt}
\bibliography{references}

\end{document}